\documentclass[twocolumn,aps,pra,longbibliography,superscriptaddress,nofootinbib,10pt]{revtex4-1}
\usepackage[latin1]{inputenc}
\usepackage{graphicx,dcolumn,bm}
\usepackage{subfigure}
\usepackage{multirow}
\usepackage{array}
\usepackage{arydshln}
\usepackage{color}
\usepackage[colorlinks,linkcolor=blue,citecolor=blue,hyperindex]{hyperref}
\usepackage{amsfonts}
\usepackage{amssymb}
\usepackage{amsmath}
\usepackage{latexsym}
\usepackage{simplewick}
\usepackage{epstopdf}
\usepackage{multirow}
\usepackage{float}
\usepackage[table]{xcolor}
\usepackage{multibib}
\usepackage{mathrsfs}
\usepackage[sans]{dsfont}

\usepackage[normalem]{ulem}

\usepackage[mathscr]{eucal}

\usepackage[normalem]{ulem}
\usepackage{epsfig}
\definecolor{Blue}{rgb}{0.0,0.0,1}
\definecolor{Red}{rgb}{1,0.0,0.0}
\definecolor{Green}{rgb}{0,0.5,0.0}
\usepackage{tikz}
\usetikzlibrary{decorations.pathmorphing}
\usetikzlibrary{arrows}
\usetikzlibrary{intersections,shapes.arrows}
\usetikzlibrary{calc}
\usetikzlibrary{quotes,angles}
\usepackage{nicefrac}
\usepackage{pgfplots}
\usepgfplotslibrary{fillbetween}
\pgfplotsset{compat=1.13,colormap={violetnew}{rgb=(0.293416, 0.0574044, 0.529412) rgb=(0.394818,0.233715,0.671945) rgb =(0.49622,0.410025,0.814477) rgb=(0.588672,0.567494,0.910066) rgb=(0.663226,0.687282,0.911765) rgb=(0.73778,0.807069,0.913465) rgb=(0.807267,0.861883,0.894034) rgb=(0.874222,0.884211,0.864039) rgb=(0.941176, 0.906538, 0.834043)}}
\usepgfplotslibrary{groupplots} 
\usepgfplotslibrary[groupplots] 
\usetikzlibrary{pgfplots.groupplots} 
\usetikzlibrary[pgfplots.groupplots] 
\usepgfplotslibrary{statistics}
\usepackage{pgfplotstable}
\tikzset{jumpdot/.style={mark=*,solid},excl/.append style={jumpdot,fill=white},incl/.append style={jumpdot,fill=black}}
\begin{document}

\title{Generalized Entropic Quantum Speed Limits}
\author{Jucelino Ferreira de Sousa}
\affiliation{Departamento de F\'{i}sica, Universidade Federal do Maranh\~{a}o, Campus Universit\'{a}rio do Bacanga, 65080-805, S\~{a}o Lu\'{i}s, Maranh\~{a}o, Brazil}
\author{Diego Paiva Pires}
\affiliation{Departamento de F\'{i}sica, Universidade Federal do Maranh\~{a}o, Campus Universit\'{a}rio do Bacanga, 65080-805, S\~{a}o Lu\'{i}s, Maranh\~{a}o, Brazil}

\begin{abstract}
We present a class of generalized entropic quantum speed limits based on $\alpha$-$z$-R\'{e}nyi relative entropy, a real-valued, contractive, two-parameter family of distinguishability measures. The quantum speed limit (QSL) falls into the class of Mandelstam-Tamm bounds, and applies to finite-dimensional quantum systems that undergo a general physical process, i.e., their effective dynamics can be modeled by unitary or nonunitary evolutions. The results cover pure or mixed, separable, and entangled probe quantum states. The QSL time depends on the smallest and largest eigenvalues of the probe and instantaneous states of the system, and its evaluation requires low computational cost. In addition, it is inversely proportional to the time-average of the Schatten speed of the instantaneous state, which in turn is fully characterized by the considered dynamics. We specialize our results to the case of unitary and nonunitary evolutions. In the former case, the QSL scales with the inverse of the energy fluctuations, while the latter depends on the Schatten $1$-norm of the rate of change of the quantum channel Kraus operators. We illustrate our findings for single-qubit and two-qubit states, and unitary and nonunitary evolutions. Our results may find applications in the study of entropic uncertainty relations, quantum metrology, and also entanglement entropies signaled by generalized entropies.
\end{abstract}

\maketitle


\section{Introduction}
\label{sec:00000000001}

The quantum speed limit (QSL) sets a fundamental threshold on the evolution time of quantum systems subjected to arbitrary physical processes~\cite{QuantInfProcess_15_3919_2016,Deffner_2017,GongHamazaki}. QSLs find applications ranging from quantum thermo\-dy\-na\-mics~\cite{Generalized_Clausius,hasegawa2023unifying,Aghion_2023}, to quantum many-body systems~\cite{Many_Body001,Many_Body002,Many_Body003,Many_Body004,Many_Body005,Many_Body006}, also including quantum metrology~\cite{giovannetti2011advances}, quantum batteries~\cite{battery001, battery002}, quantum entanglement~\cite{QSL_Entanglem002,QSL_Entanglem004}, quantum-to-classical transition~\cite{Quantum_Classical001,Quantum_Classical002,Quantum_Classical003}, optimal control theory~\cite{Optimal_Control001,Optimal_Control002}, non-Markovian dynamics~\cite{Sun2015,Meng2015,Nicolas2016,Cianciaruso2017,Teittinen_2019,Entropy_23_331_2021}, and also non-Hermitian systems~\cite{Wang_2020, Time_scaling,JPhysA_57_025302_2023,arXiv:2404.16392}. Recently, QSLs have been experimentally investigated in platforms of Nuclear Magnetic Resonance (NMR) systems~\cite{QSL_Experimental001,QSL_Experimental002}, and trapped atoms~\cite{QSL_Experimental003}.

QSLs have been originally addressed for closed quantum systems evolving between orthogonal states, reported seminaly in the works of Mandelstam and Tamm (MT)~\cite{MandelstamTamm}, ${\tau_{\text{MT}}} = \hbar\pi/(2\Delta{E})$, and Margolus and Le\-vi\-tin (ML)~\cite{MargolusLevitin}, ${\tau_{\text{ML}}} = \hbar\pi/[2(\langle{\psi_0}|{H}|{\psi_0}\rangle - {E_0})]$, where ${(\Delta{E})^2} = \langle{\psi_0}|{H^2}|{\psi_0}\rangle - {\langle{\psi_0}|{H}|{\psi_0}\rangle^2}$ is the variance of the time-independent Hamiltonian $H$, with $E_0$ being its ground state. These results paved the way for the unified QSL time ${\tau_{\text{QSL}}} = \text{max}\{{\tau_{\text{MT}}},{\tau_{\text{ML}}}\}$, proposed by Giovannetti, Lloyd and Maccone~\cite{QSL_Entanglem001,PhysRevA.67.052109}, and also Levitin and Toffoli~\cite{LevitinToffoli}. So far, QSL bounds \textit{\`{a} la} MT have been derived for time-dependent Hamiltonians and non-orthogonal, mixed, and even entangled states~\cite{PhysRevLett.65.1697}. Furthermore, a QSL was obtained for unitary dynamics that is dual to the ML bound, i.e., it scales with the inverse of the highest energy level of the system~\cite{QSL_Dual}.

The study of QSLs also covers open quantum systems. Notably, Taddei {\it et al}.~\cite{Taddei} discussed speed limits based on quantum Fisher information for a general physical process, while del Campo {\it et al}.~\cite{DelCampo} presented QSLs related to the dynamical behavior of the relative purity of quantum states. In addition, Deffner and Lutz~\cite{DeffnerLutz} proposed a class of QSLs based on the Bures angle and Schatten norms. Remarkably, a family of speed limits based on Riemannian, contractive, information-theoretic metrics on quantum state space has been proven~\cite{DiegoPires}. These ge\-ne\-ral bounds apply to unitary and nonunitary dynamics, and mixed and pure states, providing tight and attainable QSLs~\cite{PhysRevA.103.022210,NJPhys_24_055003_2022,PhysRevA.108.052207}.

The geometric approach revealed an intimate connection between QSLs and quantum state distinguishability measures. In this setting, there is a vivid in\-te\-rest in bounds based on entropic quantifiers. This includes, for e\-xam\-ple, speed limits related to von Neumann entropy~\cite{NewJPhys_24_065003}, and QSLs based on Umegaki relative entropy for unitary and nonunitary dynamics, and single and multiparticle systems~\cite{NewJPhys_24_065001_2022,arXiv:2303.07415,QSL_Entanglem003,relative_purity}. Importantly, another class of speed limits based on unified entropies has been derived for arbitrary dynamics~\cite{Unified_entropies}. It is worth mentioning a family of QSLs obtained through Petz-R\'{e}nyi relative entropy, but restricted for closed systems~\cite{bounding_generalized}. The latter provides a thermodynamic uncertainty relation for entropy production~\cite{arXiv:2402.01390}. Recently, an entropic energy-time uncertainty relation for time-independent Hamiltonians was proven based on sandwiched R\'{e}nyi conditional entropies~\cite{Renyi_Uncert_Relation}. Despite these findings, little is known about the connection between QSLs and the so-called $\alpha$-$z$-R\'{e}nyi relative entropy~\cite{ALPHAzAuden}. The latter defines a two-parameter family of relative entropies of quantum states, and it exhibits as particular cases most of the known entropic distinguishability measures~\cite{DPI-ZHANG002,Carlen_2018}. They find applications in the study of quantum coherence~\cite{PhysRevA.98.032324}, quantum channels~\cite{IEEE_65_5880_2019}, and distinguishability of random states within the framework of AdS/CFT~\cite{PRXQuantum.2.040340}.

Here, we derive a family of generalized entropic quantum speed limits based on $\alpha$-$z$-R\'{e}nyi relative entropies ($\alpha$-$z$-RREs). Our results apply for finite-dimensional quantum systems e\-vol\-ving under general physical processes. To do so, we obtain upper bounds on both non-symmetric and symmetric $\alpha$-$z$-RREs between the initial and final states of the system. Remarkably, evaluating our bounds requires minimal information about the system, namely the eigenvalues of the probe and target states, as well as the Schatten speed of the instantaneous state. The latter is fully specified in terms of the generators of the dynamics. We derive entropic QSLs for both unitary and nonunitary dynamics, which in turn fit into the class of MT bounds. We apply these results to single-qubits and two-qubits systems subjected to noisy dynamics, presenting analytical expressions and numerical simulations to support the theoretical predictions. We comment on the robustness of the bounds, also presenting a comparison with previous results in the literature.

The structure of the paper is as follows. In Sec.~\ref{sec:00000000002}, we introduce $\alpha$-$z$-RREs, and we comment on their technical properties and particular cases. In Sec.~\ref{sec:00000000003}, we derive upper bounds on $\alpha$-$z$-RREs respective to probe and instantaneous states of quantum systems undergoing general physical processes. In turn, Sec.~\ref{sec:00000000004} presents the generalized entropic quantum speed limits based on $\alpha$-$z$-RREs. We specialize the general bound to the case of unitary and nonunitary dynamics, and we discuss several particular cases. The bounds establish a two-parameter family of generalized entropic QSL times that inherit certain symmetries from $\alpha$-$z$-RREs. In Sec.~\ref{sec:00000000005}, we illustrate our findings by evaluating the QSL time for the case of a single-qubit state undergoing unitary and nonunitary evolutions, and also for a system of two-qubits coupled to independent reservoirs whose effective dynamics is described by an amplitude damping channel. Finally, in Sec.~\ref{sec:00000000006}, we summarize our conclusions.


\section{$\alpha$-$z$-R\'{e}nyi relative entropy}
\label{sec:00000000002}

In this section, we summarize the main properties of the generalized quantum relative entropies. Let $\mathcal{H}$ be the finite-dimensional Hilbert space of a given physical system, with $d = \text{dim}\mathcal{H}$. The state of the system is described by a density operator $\rho\in\mathcal{S}$, where $\mathcal{S} = \{\rho\in\mathcal{H} \mid {\rho^{\dagger}} = \rho,~\rho \geq 0,~\text{Tr}(\rho) = 1\}$ stands for the convex space of quantum states. The $\alpha$-$z$-R\'{e}nyi relative entropy ($\alpha$-$z$-RRE) for two states $\rho,\varrho\in\mathcal{S}$ is defined as~\cite{ALPHAzAuden}
\begin{equation}
\label{eq:0000000000001}
{\text{D}_{\alpha,z}}(\rho\|\varrho) =
\begin{cases}
 {\frac{1}{\alpha - 1}}\ln\left[{{{g}_{\alpha,z}}(\rho,\varrho)}\right],& \mbox{if $\text{supp}\,\rho \subseteq \text{supp}\,\varrho$} \\
+ \infty,& \mbox{otherwise}
\end{cases}
\end{equation}
with
\begin{equation}
\label{eq:0000000000002}
{g_{\alpha,z}}(\rho,\varrho) := \text{Tr}\left[\left(\varrho^{(1-\alpha)/2z}\rho^{\alpha/z}\varrho^{(1-\alpha)/2z}\right)^{z}\right] ~,
\end{equation}
for all $\alpha \in (0,1) \cup (1,+\infty)$ and $z\in\mathbb{R}^{{{}^+}}$. Here, $\text{supp}\,\omega$ stands for the support of $\omega \in \mathcal{S}$, and the logarithm is taken to base $e$. The quantity ${g_{\alpha,z}}(\rho,\varrho)$ defines the $\alpha$-$z$-relative purity, which in turn satisfies the following properties, for $z > 0$: (i) $g_{\alpha,z}(\rho,\varrho) \leq 1$, for $\alpha \in (0,1)$; (ii) ${g_{\alpha,z}}(\rho,\varrho) \geq 1$, for $\alpha \in (1,+\infty)$; (iii) $g_{\alpha,z}(\rho,\varrho) = 1$, if and only if $\rho = \varrho$, for $\alpha \in (0,1)\cup(1,+\infty)$; (iv) ${g_{1 - \alpha,z}}(\varrho,\rho) = {g_{\alpha,z}}(\rho,\varrho)$, for all $\rho,\varrho\in\mathcal{S}$, with $\alpha \in (0,1)\cup(1,+\infty)$~\cite{ALPHAzAuden,DPI_ZHANG001}.

The $\alpha$-$z$-RRE provides a general class of continuous, real-valued, two-parameter family of quantum relative entropies~\cite{ALPHAzAuden}, finding applications in quantum communication~\cite{Renyi_Communicat}, quantum thermodynamics~\cite{Renyi_Termody001,Renyi_Termody002}, and quantum coherence~\cite{Renyi_Coherence001,Renyi_Coherence002,Renyi_Coherence003}. It remains invariant under unitary transformations on the input states, i.e., ${\text{D}_{\alpha,z}}(V\rho{V^{\dagger}}\|V\varrho{V^{\dagger}}) = {\text{D}_{\alpha,z}}(\rho\|\varrho)$, with $V{V^{\dagger}} = {V^{\dagger}}V = \mathbb{I}$. Furthermore, for all $\rho,\varrho,\varpi,\varsigma\in\mathcal{S}$, one finds that ${\text{D}_{\alpha,z}}(\rho\otimes\varpi\|\varrho\otimes\varsigma) = {\text{D}_{\alpha,z}}(\rho\|\varrho) + {\text{D}_{\alpha,z}}(\varpi\|\varsigma)$. Note that $\alpha$-$z$-RRE satisfies the data-processing ine\-qua\-li\-ty (DPI), i.e., ${\text{D}_{\alpha,z}}(\mathcal{E}(\rho)\|\mathcal{E}(\varrho))\leq{\text{D}_{\alpha,z}}(\rho\|\varrho)$, where $\mathcal{E}(\bullet)$ is a completely-positive and trace-preserving (CPTP) map, if and only if one of the fol\-lowing conditions holds: (i) $\alpha \in (0,1)$ and $z \geq \text{max}\{\alpha,1 - \alpha\}$; (ii) $\alpha \in (1,2)$ and $z \in [\alpha/2,\alpha]$; (iii) $\alpha \in [2,\infty)$ and $z \in [\alpha -1,\alpha]$~\cite{DPI_ZHANG001,DPI-ZHANG002,Carlen_2018}. The DPI ensures that local operations on the quantum state cannot increase the information content. Hence, satisfying the DPI is an important criterion for general distinguisha\-bi\-li\-ty measures of quantum states in the scope of quantum information science.

On the one hand, for $z = 1$, $\alpha$-$z$-RRE reduces to the Petz-R\'{e}nyi relative entropy (PRRE), defined as ${\text{D}_{\alpha,1}}(\rho\|\varrho) = {\text{R}_{\alpha}}(\rho\|\varrho) := {(\alpha - 1)^{-1}}\ln[\text{Tr}({\rho^{\alpha}}{\varrho^{1 - \alpha}})]$~\cite{PETZ}. In turn, for all $\rho,\varrho \in \mathcal{S}$, PRRE is (i) nonnegative, i.e., ${\text{R}_{\alpha}}(\rho\|\varrho) \geq 0$; and (ii) monotonic, ${\text{R}_{\alpha}}(\mathcal{E}(\rho)\|\mathcal{E}(\varrho)) \leq {\text{R}_{\alpha}}(\rho\|\varrho)$, for $\alpha \in (0,1)\cup(1,2)$~\cite{Renyi_monotonic}. On the other hand, for $z = \alpha$, we have that $\alpha$-$z$-RRE recovers the so-called sandwiched R\'{e}nyi relative entropy (SRRE), also known as $\alpha$-quantum R\'{e}nyi divergence, defined as $\text{D}_{\alpha,\alpha}(\rho\|\varrho) = \widetilde{\text{D}}_{\alpha}(\rho\|\varrho) := {(\alpha - 1)^{-1}}\ln[\text{Tr}({\varrho^{(1 - \alpha)/{2\alpha}}}\rho{\varrho^{(1-\alpha)/{2\alpha}}})^{\alpha}]$~\cite{sandwiched001,sandwiched002}. We note that SRRE is monotonic, $\widetilde{\text{D}}_{\alpha}(\mathcal{E}(\rho)\|\mathcal{E}(\varrho))\leq\widetilde{\text{D}}_{\alpha}(\rho\|\varrho)$, for $\alpha \in [1/2,\infty)$~\cite{Monotonicity_Sandwiched}. It has been ve\-ri\-fied that SRRE is upper bounded by PRRE, i.e., $\widetilde{\text{D}}_{\alpha}(\rho\|\varrho) \leq {\text{R}_{\alpha}}(\rho\|\varrho)$~\cite{JPhysAMathTheor_47_045304,sandwiched002}. In particular, for $\alpha = 1/2$, SRRE reduces to the min-relative entropy, $\widetilde{\text{D}}_{1/2}(\rho\|\varrho) = {\text{D}_{\text{min}}}(\rho\|\varrho) := - 2\ln[F(\rho,\varrho)]$, where $F(\rho,\varrho) = \text{Tr}\left(\sqrt{\sqrt{\rho}{\varrho}\sqrt{\rho}} \, \right)$ is the Uhlmann fidelity. Next, for $\alpha \rightarrow \infty$, SRRE reduces to the max-relative entropy, $\widetilde{\text{D}}_{\alpha \rightarrow \infty}(\rho\|\varrho) = {\text{D}_{\text{max}}}(\rho\|\varrho) := \text{inf}\{\gamma\in\mathbb{R} \, | \, \rho \leq \exp(\gamma)\varrho\}$~\cite{min_max_relat, sandwiched001}. It is noteworthy that, for $\alpha \rightarrow 1$, both PRRE and SRRE collapse into the Umegaki's relative entropy (URE), $\text{S}(\rho\|\varrho) = -S(\rho) - \text{Tr}(\rho\ln\varrho)$, where $S(\rho) = - \text{Tr}(\rho\ln\rho)$ is the von Neumann entropy~\cite{RelEntr02}. In turn, URE quantifies the amount of quantum coherence and entanglement~\cite{Coher001_Relat_Entrop, Coher002_Relat_Entrop,Entang_Relat_Entrop}, also useful in quantum thermodynamics~\cite{Entropy_Production001,Entropy_Production002,Entropy_Production003}, and asymmetry theory~\cite{marvian2014extending}.

Overall, $\alpha$-$z$-RRE is skew symmetric, i.e., $(1 - \alpha){\text{D}_{\alpha,z}}(\rho\|\varrho) = \alpha{\text{D}_{1 - \alpha,z}}(\varrho\|\rho)$, with $\alpha \in (0,1)\cup(1,+\infty)$ and $z > 0$~\cite{ALPHAzAuden}. In particular, we have that $\alpha$-$z$-RRE is symmetric for $\alpha = 1/2$, i.e., ${\text{D}_{1/2,z}}({\rho}\|{\varrho}) = {\text{D}_{1/2,z}}({\varrho}\|{\rho})$. Importantly, with the exception of this case, it turns out that $\alpha$-$z$-RRE is asymmetric with respect to the exchange of the input states, i.e., ${\text{D}_{\alpha,z}}({\rho}\|{\varrho}) \neq {\text{D}_{\alpha,z}}({\varrho}\|{\rho})$. The fact that $\alpha$-$z$-RRE is asymmetric indicates that such a quantifier does not fulfill the triangle inequality. The last two points suggest that $\alpha$-$z$-RRE fails to be a metric in the strict sense of a proper distance embedded in a me\-tric space. To clarify, by genuine metrics we refer to the class of information-theoretic quantifiers that are non\-nega\-ti\-ve, symmetric with respect to the input states, and also satisfy the triangle inequality~\cite{Ingemar_Bengtsson_Zyczkowski}. It is worth noting that Umegaki's relative entropy mentioned above is also not a metric in the space of quantum states. However, it is clear that such an entropic quantifier stands out as a useful distinguishability measure of quantum states, and plays an important role in several branches of quantum information science~\cite{Coher001_Relat_Entrop, Coher002_Relat_Entrop,Entang_Relat_Entrop,Entropy_Production001,Entropy_Production002,Entropy_Production003,marvian2014extending}. We note that Petz-R\'{e}nyi and sandwiched R\'{e}nyi relative entropies have been used in the study of infinite families of second laws of thermodynamics~\cite{Renyi_Termody001}. We note, however, that both are not symmetrical and therefore do not define faithful metrics.

In particular, the lack of symmetry of $\alpha$-$z$-RRE can be mitigated by introducing an \textit{ad hoc} symmetrized version of such a measure defined as follows
\begin{equation}
\label{eq:0000000000003}
{\text{D}_{\alpha,z}}(\rho:\varrho) := {\text{D}_{\alpha,z}}(\rho\|\varrho) + {\text{D}_{\alpha,z}}(\varrho\|\rho) ~.
\end{equation}
In this case, one readily finds that ${\text{D}_{\alpha,z}}({\rho}:{\varrho}) = {\text{D}_{\alpha,z}}({\varrho}:{\rho})$, for all $z > 0$ and $\alpha \in (0,1)\cup(1,+\infty)$. Furthermore, Eq.~\eqref{eq:0000000000003} shows that the symmetrized $\alpha$-$z$-RRE is skew symmetric with respect to the para\-me\-ter $\alpha \in (0,1)\cup(1,+\infty)$, i.e., $(1 - \alpha){\text{D}_{\alpha,z}}(\rho:\varrho) = \alpha{\text{D}_{1 - \alpha,z}}(\rho:\varrho)$, for all $z > 0$. We emphasize that, to our knowledge, there is no evidence that ${\text{D}_{\alpha,z}}(\rho:\varrho)$ satisfies the triangle inequality. However, there is no detriment to the discussion and results addressed throughout the manuscript. We investigate $\alpha$-$z$-RRE as a useful figure of merit in the study of quantum speed limits, thus providing a family of entropic QSLs exploiting the sensitivity of $\alpha$-$z$-RRE to distinguish two neighboring quantum states. In the following, we present a class of upper bounds on $\alpha$-$z$-RREs for finite-dimensional quantum systems undergoing arbitrary physical processes.


\section{Bounding ${\alpha}$-$z$-RRE}
\label{sec:00000000003}

We consider a quantum system undergoing a general, time-dependent physical process. The instantaneous state of the system is given by ${\rho_t} = {\mathcal{E}_t}({\rho_0})$, with $t \in [0,\tau]$, where ${\mathcal{E}_t}(\bullet)$ is a physical evolution map, and ${\rho_0}\in\mathcal{S}$ defines the initial state of the system. Hereafter, we focus on the range of parameters $0 < \alpha < 1$ and $1 \geq z \geq \text{max}\{\alpha,1 - \alpha\}$ where $\alpha$-$z$-RRE satisfies the data processing inequality, and we set $\hbar = 1$. In this setting, we find that the $\alpha$-$z$-R\'{e}nyi with respect to states $\rho_0$ and ${\rho_{\tau}}$ is upper bounded as follows [see Appendix~\ref{sec:0000000000B01}]
\begin{equation}
\label{eq:0000000000004}
{\text{D}_{\alpha,z}}({\rho_{\tau}}\|{\rho_0}) \leq \frac{\alpha \, {h_{\alpha,z}}({\rho_0})}{\left|1 - \alpha\right|}{\int_0^{\tau}} dt \, {\left[{k_{\text{min}}}({\rho_t})\right]^{\alpha - 1}}{\left\|\frac{d{\rho_t}}{dt}\right\|_1} ~,
\end{equation}
where ${\|X\|_1} := \text{Tr}(\sqrt{{X^{\dagger}}X})$, with
\begin{equation}
\label{eq:0000000000005}
{h_{\alpha,z}}({\rho_0}) := \frac{\left({k_{\text{max}}}({\rho_0})\left[{k_{\text{min}}}({\rho_0})\right]^{z - 1}\right)^{\frac{1 - \alpha}{z}}}{\left|1 + (1 - \alpha)\ln[k_{\text{min}}({\rho_0})]\right|} ~,
\end{equation}
while $k_{\text{min/max}}(X)$ sets the smallest/largest eigenvalue of the state $X \in \mathcal{S}$. We note that another class of upper bounds on $\alpha$-$z$-RREs can be derived when the order of the input states ${\rho_0}$ and ${\rho_{\tau}}$ in Eq.~\eqref{eq:0000000000004} is swapped, thus yielding [see Appendix~\ref{sec:0000000000B02}]
\begin{equation}
\label{eq:0000000000006}
{\text{D}_{\alpha,z}}({\rho_0}\|{\rho_{\tau}}) \leq {h_{1 - \alpha, z}}({\rho_0}){\int_0^{\tau}} dt \, {\left[{k_{\text{min}}}({\rho_t})\right]^{-\alpha}}{\left\|\frac{d{\rho_t}}{dt}\right\|_1} ~.
\end{equation}
Importantly, using the fact that $\alpha$-$z$-RRE is skew symmetric, it follows that the bounds in Eqs.~\eqref{eq:0000000000004} and~\eqref{eq:0000000000006} can be mapped into each other through the transformation $\alpha \rightarrow 1 - \alpha$, for all $z > 0$. It is worth noting that an upper bound for the symmetrized $\alpha$-$z$-RRE in Eq.~\eqref{eq:0000000000003} can be obtained in a similar fashion. In fact, by com\-bi\-ning the non-symmetric bounds in Eqs.~\eqref{eq:0000000000004} and~\eqref{eq:0000000000006}, one finds [see Appendix~\ref{sec:0000000000B03}]
\begin{align}
\label{eq:0000000000007}
&{\text{D}_{\alpha,z}}({\rho_0}:{\rho_{\tau}}) \leq \frac{\alpha \, {h_{\alpha,z}}({\rho_0})}{\left|1 - \alpha\right|}{\int_0^{\tau}} dt \, {\left[{k_{\text{min}}}({\rho_t})\right]^{\alpha - 1}}{\left\|\frac{d{\rho_t}}{dt}\right\|_1} \nonumber\\
&+ {h_{1 - \alpha, z}}({\rho_0}){\int_0^{\tau}} dt \, {\left[{k_{\text{min}}}({\rho_t})\right]^{-\alpha}}{\left\|\frac{d{\rho_t}}{dt}\right\|_1} ~,
\end{align}
with the auxiliary, time-independent functions ${h_{\alpha,z}}({\rho_0})$ and ${h_{1 - \alpha,z}}({\rho_0})$ given by Eq.~\eqref{eq:0000000000005}. We point out that the inequality in Eq.~\eqref{eq:0000000000007} remains invariant under the mapping $\alpha \rightarrow 1 - \alpha$. This comes from the skew symmetry $(1 - \alpha){\text{D}_{\alpha,z}}({\rho_0}:{\rho_{\tau}}) = \alpha{\text{D}_{1 - \alpha,z}}({\rho_0}:{\rho_{\tau}})$ exhibited by the symmetrized $\alpha$-$z$-RRE.

Importantly, Eqs.~\eqref{eq:0000000000004},~\eqref{eq:0000000000006}, and~\eqref{eq:0000000000007} display the main results of the paper. The $\alpha$-$z$-RREs in the left-hand side of these bounds characterize the distinguishability between the states ${\rho_0}$ and ${\rho_{\tau}}$, thus assigning the geo\-me\-tric notion of distance measure between initial and final states of the system. In fact, the closer the two states are, the smaller their $\alpha$-$z$-RREs will be. The bounds also depend on the Schatten speed ${\|d{\rho_t}/dt\|_1}$, i.e., the quantum speed that is induced by the dynamical physical process~\cite{StatistSpeedAugusto,Deffner_Wigner_Space}. We point out that the Schatten speed is fully characterized by the generator that governs the evolution of the quantum system.

We find that the right-hand sides of Eqs.~\eqref{eq:0000000000004},~\eqref{eq:0000000000006}, and~\eqref{eq:0000000000007} depend on the au\-xi\-liary function ${h_{\alpha,z}}({\rho_0})$ defined in Eq.~\eqref{eq:0000000000005}. The latter is a real-valued, time-independent, two-parameter function, that depends on both the smallest ${k_{\text{min}}}({\rho_0})$ and largest ${k_{\text{max}}}({\rho_0})$ eigen\-va\-lues of the initial state. In addition, the bounds also depend on the smallest eigenvalue ${k_{\text{min}}}({\rho_t})$ of the evolved state of the system. Notably, evaluating these bounds is expected to require low computational cost. This can be advantageous, for e\-xam\-ple, in studying the dynamical behavior of $\alpha$-$z$-RREs for higher-dimensional quantum systems, where pre\-dic\-ting the full spectrum of the density matrices is typically a difficult task.

In particular, for $\alpha = 1/2$, Eqs.~\eqref{eq:0000000000004},~\eqref{eq:0000000000006} and~\eqref{eq:0000000000007} identically collapses into the bound as follows
\begin{equation}
\label{eq:0000000000008}
{\text{D}_{1/2,z}}({\rho_0}\|{\rho_{\tau}}) \leq {h_{1/2,z}}({\rho_0}) {\int_0^{\tau}} dt \, \frac{{\|d{\rho_t}/dt\|_1}}{\sqrt{{k_{\text{min}}}({\rho_t})}} ~,
\end{equation}
where we have used that ${\text{D}_{1/2,z}}({\rho_0}\|{\rho_{\tau}}) = {\text{D}_{1/2,z}}({\rho_{\tau}}\|{\rho_0})$ is symmetric for $\alpha = 1/2$, for all $1/2 < z \leq 1$. In turn, by fixing $z = 1$, Eq.~\eqref{eq:0000000000008} provides an upper bound to the quantum affinity related to states $\rho_{\tau}$ and $\rho_0$, which can be useful in the study of quantum metrology.

To investigate the tightness of the bound on $\alpha$-$z$-RRE in Eq.~\eqref{eq:0000000000007}, we introduce the figure of merit as follows
\begin{equation}
\label{eq:0000000000008001}
{\delta^{\text{BOUND}}_{\alpha,z}}(\tau) := 1 - \frac{{\text{D}_{\alpha,z}}({\rho_0}:{\rho_{\tau}})}{\frac{1}{|1 - \alpha|}{\int_0^{\tau}} dt \, {\Phi_{\alpha,z}}({\rho_0},{\rho_t}) {\left\|\frac{d{\rho_t}}{dt}\right\|_1}} ~,
\end{equation}
where we define the time-dependent auxiliary function
\begin{align}
\label{eq:0000000000012}
&{\Phi_{\alpha,z}}({\rho_0},{\rho_t}) := \alpha \, {h_{\alpha,z}}({\rho_0}){\left[{k_{\text{min}}}({\rho_t})\right]^{\alpha - 1}} \nonumber\\
&+ |1 - \alpha| \, {h_{1 - \alpha,z}}({\rho_0}){\left[{k_{\text{min}}}({\rho_t})\right]^{-\alpha}} ~.
\end{align}
It is worth to note that Eq.~\eqref{eq:0000000000012} fulfills the symmetry ${\Phi_{\alpha,z}}({\rho_0},{\rho_t}) = {\Phi_{1 - \alpha,z}}({\rho_0},{\rho_t})$, for all $\alpha \in (0,1)$. The smaller the relative error in Eq.~\eqref{eq:0000000000008001}, the tighter the bound on $\alpha$-$z$-RRE. The relative error captures the deviation of the $\alpha$-$z$-RRE with respect to the average quantum speed. The bound saturates when the $\alpha$-$z$-RRE coincides with the time-average of the product between the weight function ${\Phi_{\alpha,z}}({\rho_0},{\rho_t})$ and the quantum speed $\|d{\rho_t}/dt\|_1$ induced by the overall dynamics.


\section{Generalized Entropic QSL}
\label{sec:00000000004}

In this section, we establish families of entropic QSLs based on $\alpha$-$z$-RREs. The main idea is to derive lower bounds for the evolution time of quantum systems subjected to arbitrary physical processes that depend on the distinguishability of states signaled by general relative entropies. Based on Eqs.~\eqref{eq:0000000000004},~\eqref{eq:0000000000006}, and~\eqref{eq:0000000000007}, the time required for a quantum system to reach the state ${\rho_{\tau}}$, star\-ting from the initial state $\rho_0$, satisfies the lower bound $\tau \geq \tau^{\text{QSL}}_{\alpha,z}$, with the generalized entropic QSL time given by
\begin{equation}
\label{eq:0000000000009}
{\tau^{\text{QSL}}_{\alpha,z}} := \text{max}\{{\tau_{\alpha,z}}({\rho_{\tau}}\|{\rho_0}), {\tau_{\alpha,z}}({\rho_0}\|{\rho_{\tau}}),{\tau_{\alpha,z}}({\rho_0}:{\rho_{\tau}})\} ~,
\end{equation}
with
\begin{equation}
\label{eq:0000000000010}
{\tau_{\alpha,z}}({\rho_{\tau}}\|{\rho_0}) := \frac{|1 - \alpha| \, {\text{D}_{\alpha,z}}({\rho_{\tau}}\|{\rho_0})}{\alpha \, {h_{\alpha,z}}({\rho_0})\langle\langle {\left[{k_{\text{min}}}({\rho_t})\right]^{\alpha - 1}} {\|d{\rho_t}/dt\|_1} \rangle\rangle_{\tau}} ~,
\end{equation}
where ${\langle\langle \bullet \rangle\rangle_{\tau}} := {\tau^{-1}}{\int_0^{\tau}}dt\,\bullet$ stands for the time-average, while the QSL related to the symmetrized $\alpha$-$z$-RREs reads
\begin{equation}
\label{eq:0000000000011}
{\tau_{\alpha,z}}({\rho_0}:{\rho_{\tau}}) := \frac{|1 - \alpha| \, {\text{D}_{\alpha,z}}({\rho_0}:{\rho_{\tau}})}{\langle\langle{\Phi_{\alpha,z}}({\rho_0},{\rho_t}){\|d{\rho_t}/dt\|_1} \rangle\rangle_{\tau}} ~,
\end{equation}
with the time-dependent auxiliary function ${\Phi_{\alpha,z}}({\rho_0},{\rho_t})$ defined in Eq.~\eqref{eq:0000000000012}. We note that ${\tau_{\alpha,z}}({\rho_0}\|{\rho_{\tau}}) = {\tau_{1 - \alpha, z}}({\rho_{\tau}}\|{\rho_0})$, for all $0 < \alpha < 1$ and $1 \geq z \geq \text{max}\{\alpha, 1 - \alpha\}$. In addition, one finds that Eq.~\eqref{eq:0000000000011} is invariant under the mapping $\alpha \rightarrow 1 - \alpha$, i.e., ${\tau_{\alpha,z}}({\rho_0}:{\rho_{\tau}}) = {\tau_{1 - \alpha,z}}({\rho_0}:{\rho_{\tau}})$. Therefore, putting these results together, we have that the generalized QSL time in Eq.~\eqref{eq:0000000000009} sa\-tis\-fies the symmetry ${\tau^{\text{QSL}}_{\alpha,z}} = {\tau^{\text{QSL}}_{1 - \alpha,z}}$, for all ${\rho_0},{\rho_{\tau}}\in\mathcal{S}$.

The quantity ${\tau^{\text{QSL}}_{\alpha,z}}$ in Eq.~\eqref{eq:0000000000009} sets a two-parameter fa\-mi\-ly of entropic quantum speed limits, and it stands as the second main result of the paper. It follows that ${\tau^{\text{QSL}}_{\alpha,z}}$ covers both unitary and nonunitary physical process, pure or mixed states, and entangled or separable states. Overall, it depends on the distinguishability of the initial and final states of the quantum system that is captured by the $\alpha$-$z$-RRE. Notably, Eq.~\eqref{eq:0000000000009} requires minimal information about the spectrum of the density matrix, namely the smallest and largest eigenvalues of the probe and target states of the dynamics. The QSL time scales with the inverse of the time-average of the Schatten speed, which in turn signals the quantum fluctuations of the generator of the dy\-na\-mi\-cal process.
\begin{figure}[!t]
\begin{center}
\includegraphics[scale=1.1]{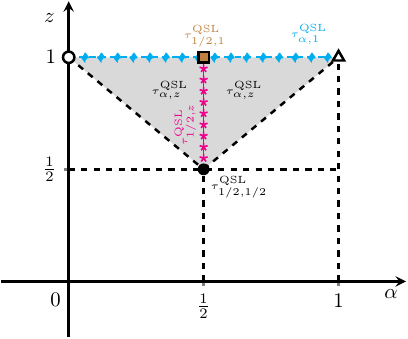}
\caption{(Color online) Phase diagram for the generalized entropic quantum speed limit, $\tau_{\alpha,z}^{\text{QSL}}$ [see Eq.~\eqref{eq:0000000000009}]. The gray shaded area indicates the region in which the $\alpha$-$z$-QSL applies, namely $0 < \alpha < 1$ and $1 \geq z \geq \text{max}\{\alpha,1 - \alpha\}$. It also restricts the pairs $(\alpha,z)$ for which $\alpha$-$z$-QSL satisfies the data processing inequality. For $z = 1$, one finds that the speed limit reduces to $\tau_{\alpha,1}^{\text{QSL}}$ (cyan cross marks), which in turn is related to the Petz-R\'{e}nyi relative entropy. For $\alpha = 1/2$, one gets the QSL time ${\tau_{1/2,z}^{\text{QSL}}}$ (magenta star marks). In particular, for $\alpha = z = 1/2$, the entropic speed limit ${\tau_{1/2,1/2}^{\text{QSL}}}$ depends on the Uhlmann fidelity (black dot mark). However, for $\alpha = 1/2$ and $z = 1$, the QSL time ${\tau_{1/2,1}^{\text{QSL}}}$ is related to the quantum affinity (brown square mark). We emphasize that the bounds do not apply for $\alpha = 0$ (white circle mark) and $\alpha = 1$ (white triangle mark).}
\label{FIG000001}
\end{center}
\end{figure}

Next, we comment on the entropic speed limit for some particular cases of $\alpha$ and $z$. These results are re\-pre\-sen\-ted schematically in the phase diagram shown in Fig.~\ref{FIG000001}, which also shows the region in which the entropic speed limit applies. On the one hand, for $z = 1$, Eq.~\eqref{eq:0000000000009} [see also Eqs.~\eqref{eq:0000000000010}, and~\eqref{eq:0000000000011}] implies the QSL time $\tau_{\alpha,1}^{\text{QSL}}$, which in turn is related to the Petz-R\'{e}nyi relative entropy. In particular, using the Pinsker inequality, it is known that PRRE is bounded from below by trace distance. In this case, Eqs.~\eqref{eq:0000000000009},~\eqref{eq:0000000000010} and~\eqref{eq:0000000000011} become similar to the QSLs addressed in Refs.~\cite{Deffner_Wigner_Space,StatistSpeedAugusto,NewJPhys_21_013006_2019,PhysRevLett.126.010601,NewJPhys_24_095004_2022,arXiv_231200533,arXiv:2401.01746}. On the other hand, for $\alpha = 1/2$ and $z \geq 1/2$, we find the QSL time ${\tau_{1/2,z}^{\text{QSL}}}$. We remind the reader that, for $\alpha = 1/2$, the generalized relative entropy is symmetric with respect to the input states, thus im\-plying that the latter QSL time exhibits the symmetry as follows
\begin{equation}
\label{eq:0000000000013}
{\tau_{1/2,z}^{\text{QSL}}} = {\tau_{1/2,z}}({\rho_0}\|{\rho_{\tau}}) = {\tau_{1/2,z}}({\rho_{\tau}}\|{\rho_0}) = {\tau_{1/2,z}}({\rho_0}:{\rho_{\tau}}) ~,
\end{equation}
for all $z > 1/2$. In turn, for $\alpha = z = 1/2$, the entropic speed limit depends on the Uhlmann fidelity. However, for $\alpha = 1/2$ and $z = 1$, the QSL time is related to the quantum affinity.

It is worthwhile to notice that the generalized entropic QSL in Eq.~\eqref{eq:0000000000009} [see also Eqs.~\eqref{eq:0000000000010}, and~\eqref{eq:0000000000011}] is incompatible with the speed limit discussed in Ref.~\cite{Unified_entropies}. In fact, the latter result applies to unified entropies, i.e., a two-parameter entropic family that has as input only the instantaneous state of the quantum system. In turn, we have that the generalized QSL is based on the dynamical behavior of $\alpha$-$z$-RRE, i.e., it takes advantage of the distinguishability between probe and target states of the dynamics. In this sense, the two results do not appear to be equivalent to each other.

To investigate the tightness of the entropic speed limit time in Eq.~\eqref{eq:0000000000009}, we introduce the relative error 
\begin{align}
\label{eq:000000000001302}
&{\delta^{\text{QSL}}_{\alpha,z}}(\tau) := \nonumber\\
& 1 - \frac{\text{max}\{{\tau_{\alpha,z}}({\rho_{\tau}}\|{\rho_0}), {\tau_{\alpha,z}}({\rho_0}\|{\rho_{\tau}}),{\tau_{\alpha,z}}({\rho_0}:{\rho_{\tau}})\}}{\tau} ~.
\end{align}
The smaller the relative error in Eq.~\eqref{eq:000000000001302}, the tighter the entropic speed limit time. In the following, we will address these QSL times by specifying the dynamics of the quantum system.


\subsection{Closed quantum systems}
\label{sec:00000000004A}

Let ${\mathcal{E}_t}(\bullet) = {U_t}\bullet{U_t^{\dagger}}$ be the unitary evolution that drives the physical process, with ${U_t} = {e^{- i t H}}$ being the evolution operator, and $H$ sets a time-independent Hamiltonian. In this case, the Schatten speed becomes ${\|d{\rho_t}/dt\|_1} = {\|(-i)[H,{\rho_0}]\|_1}$. In addition, it has been proved that $ {\|(-i)[H,{\rho_0}]\|_1} \leq 2 \sqrt{\mathcal{Q}({\rho_t})}$, where $\mathcal{Q}({\rho_t}) = (1/4)\text{Tr}({\rho_t}{L_t^2})$ is the quantum Fisher information (QFI), with ${L_t}$ being the symmetric logarithmic derivative~\cite{boundFisher}. In turn, QFI fulfills the upper bound $\mathcal{Q}({\rho_t}) \leq {(\Delta{H})^2}$, where ${(\Delta{H})^2} = \text{Tr}({\rho_0}{H^2}) - {[\text{Tr}({\rho_0}H)]^2}$ is the variance of $H$~\cite{SYu_1302.5311,PhysRevA.87.032324,GToth_1701.07461}. Therefore, one concludes that the Schatten speed satisfies the upper bound ${\|d{\rho_t}/dt\|_1} \leq 2\Delta{H}$, which is saturated by initial pure states. We also note that ${k_{\text{min}/\text{max}}}({\rho_t}) = {k_{\text{min}/\text{max}}}({\rho_0})$, since the spectrum of the density matrix remains invariant under unitary evolutions. 

In this setting, the generalized entropic QSL time for unitary evolutions (UQSL), i.e., closed quantum systems, reads as
\begin{equation}
\label{eq:0000000000014}
{\tau^{\text{UQSL}}_{\alpha,z}} := \text{max}\{{\tau'_{\alpha,z}}({\rho_{\tau}}\|{\rho_0}), {\tau'_{\alpha,z}}({\rho_0}\|{\rho_{\tau}}),{\tau'_{\alpha,z}}({\rho_0}:{\rho_{\tau}})\} ~,
\end{equation}
where Eq.~\eqref{eq:0000000000010} becomes
\begin{equation}
\label{eq:0000000000015}
{\tau'_{\alpha,z}}({\rho_{\tau}}\|{\rho_0}) := \frac{|1 - \alpha| \, {\text{D}_{\alpha,z}}({\rho_{\tau}}\|{\rho_0})}{2 \alpha \, {h_{\alpha,z}}({\rho_0}) {\left[{k_{\text{min}}}({\rho_0})\right]^{\alpha - 1}} \Delta{H}} ~,
\end{equation}
with ${\tau'_{\alpha,z}}({\rho_0}\|{\rho_{\tau}}) = {\tau'_{1 - \alpha, z}}({\rho_{\tau}}\|{\rho_0})$, while Eq.~\eqref{eq:0000000000011} implies the lower bound
\begin{align}
\label{eq:0000000000016}
&{\tau'_{\alpha,z}}({\rho_0}:{\rho_{\tau}}) := \nonumber\\
& \frac{|1 - \alpha| \, {\left[{k_{\text{min}}}({\rho_0})\right]^{\alpha}} \, {\text{D}_{\alpha,z}}({\rho_0}:{\rho_{\tau}})}{2\{\alpha \, {h_{\alpha,z}}({\rho_0}){\left[{k_{\text{min}}}({\rho_0})\right]^{2\alpha - 1}} + |1 - \alpha| \, {h_{1 - \alpha,z}}({\rho_0})\}\Delta{H} } ~.
\end{align}
Notably, Eqs.~\eqref{eq:0000000000015} and~\eqref{eq:0000000000016} are inversely proportional to the variance of the Hamiltonian of the closed system, which implies that the entropic QSLs fall into the Mandelstam-Tamm class of speed limits. The more distinguishable the two states $\rho_0$ and $\rho_{\tau}$ are, the larger the lower bound on the evolution time.

Next, we specialize the result in Eq.~\eqref{eq:0000000000015} [see also Eq.~\eqref{eq:0000000000016}]. On the one hand, for $z = 1$, one obtains the QSL time as follows
\begin{align}
\label{eq:0000000000017}
&{\tau'_{\alpha,1}}({\rho_{\tau}}\|{\rho_0}) = \nonumber\\
& \frac{|1 - \alpha| \, |1 + (1 - \alpha)\ln[k_{\text{min}}({\rho_0})]| \, {\text{R}_{\alpha}}({\rho_{\tau}}\|{\rho_0})}{2 \, \alpha \, {[{k_{\text{max}}}({\rho_0})]^{1 - \alpha}}{[{k_{\text{min}}}({\rho_0})]^{\alpha - 1}}\Delta{H}} ~,
\end{align}
where ${\text{R}_{\alpha}}({\rho_{\tau}}\|{\rho_0})$ is the Petz-R\'{e}nyi relative entropy [see Sec.~\ref{sec:00000000002}]. Note that the latter bound is similar to the QSLs addressed in Ref.~\cite{bounding_generalized} for the case of time-independent Hamiltonians. We note, however, that such a result was based on Schatten $2$-norm that may provide loose bounds compare to the results discussed here. On the other hand, for $\alpha = z = 1/2$, Eq.~\eqref{eq:0000000000015} implies the positive, real-valued lower bound 
\begin{align}
\label{eq:0000000000018}
&{\tau'_{1/2,1/2}}({\rho_{\tau}}\|{\rho_0}) = \nonumber\\
&- \frac{|2 + \ln[k_{\text{min}}({\rho_0})]| {k_{\text{min}}}({\rho_0}) \ln[F({\rho_{\tau}},{\rho_0})]}{2 \, {k_{\text{max}}}({\rho_0}) \, \Delta{H}} ~,
\end{align}
where $F({\rho_{\tau}},{\rho_0}) = \text{Tr}\left(\sqrt{\sqrt{{\rho_{\tau}}}{\rho_0}\sqrt{{\rho_{\tau}}}} \, \right)$ is the Uhlmann fidelity, with $0 \leq F({\rho_{\tau}},{\rho_0}) \leq 1$. We note, however, that the bound in Eq.~\eqref{eq:0000000000018} is loose for the case of indistinguishable or orthogonal quantum states, since it becomes trivially vanishing or approaches large values, respectively. In the former case, for example, Eq.~\eqref{eq:0000000000018} becomes ${\tau'_{1/2,1/2}}({\rho_{\tau}}\|{\rho_0}) \approx |2 + \ln[k_{\text{min}}({\rho_0})]| {k_{\text{min}}}({\rho_0})[1 - F({\rho_{\tau}},{\rho_0})]/[2 \, {k_{\text{max}}}({\rho_0}) \, \Delta{H}]$, for $F({\rho_{\tau}},{\rho_0}) \approx 1$. We note that, for time-independent Hamiltonians and mixed states, the MT bound approaches ${\tau_{\text{MT}}} \approx \sqrt{2[1 - F({\rho_{\tau}},{\rho_0})]}/\Delta{H}$ whenever $F({\rho_{\tau}},{\rho_0}) \approx 1$~\cite{MandelstamTamm,Deffner_2017}.


\subsection{Open quantum systems}
\label{sec:00000000004B}

Next, let $\mathcal{E}_{t}(\bullet) = {\sum_l}\,{K_l}\bullet{K^{\dagger}_l}$ be the CPTP map that governs the nonunitary evolution of a given open quantum system, where ${\{{K_l}\}_{l = 1,\ldots,m}}$ is the set of time-dependent Kraus operators, with ${\sum_l}\,{K^{\dagger}_l}{K_l} = \mathbb{I}$~\cite{NIELSEN}. In this case, one verifies that the Schatten speed fulfills the upper bound ${\|d{\rho_t}/dt\|_1} \leq 2 \, {\sum_l}\,{\|{K_l}{\rho_0}(d{K_l^{\dagger}}/dt)\|_1}$, where we have applied the triangular inequality ${\|{\sum_l}\,{\mathcal{O}_l}\|_1} \leq {\sum_l} {\|{\mathcal{O}_l}\|_1}$, and we also used the property ${\|\mathcal{O}^{\dagger}\|_1} = {\|\mathcal{O}\|_1}$. Hence, from Eq.~\eqref{eq:0000000000009} [see also Eqs.~\eqref{eq:0000000000010} and~\eqref{eq:0000000000011}], we obtain the generalized entropic quantum speed limit for nonunitary evolutions (NUQSL), i.e., open quantum systems, as follows
\begin{equation}
\label{eq:0000000000019}
{\tau^{\text{NUQSL}}_{\alpha,z}} := \text{max}\{{\tau''_{\alpha,z}}({\rho_{\tau}}\|{\rho_0}), {\tau''_{\alpha,z}}({\rho_0}\|{\rho_{\tau}}),{\tau''_{\alpha,z}}({\rho_0}:{\rho_{\tau}})\} ~,
\end{equation}
with
\begin{align}
\label{eq:0000000000020}
&{\tau''_{\alpha,z}}({\rho_{\tau}}\|{\rho_0}) = \nonumber\\
& \frac{|1 - \alpha| \, {\text{D}_{\alpha,z}}({\rho_{\tau}}\|{\rho_0})}{2 \alpha \, {h_{\alpha,z}}({\rho_0}){\sum_l}\langle\langle {\left[{k_{\text{min}}}({\rho_t})\right]^{\alpha - 1}}{\|{K_l}{\rho_0}(d{K_l^{\dagger}}/dt)\|_1} \rangle\rangle_{\tau}} ~,
\end{align}
where ${\tau''_{\alpha,z}}({\rho_0}\|{\rho_{\tau}}) = {\tau''_{1 - \alpha, z}}({\rho_{\tau}}\|{\rho_0})$, and 
\begin{equation}
\label{eq:0000000000021}
{\tau''_{\alpha,z}}({\rho_0}:{\rho_{\tau}}) := \frac{|1 - \alpha| \, {\text{D}_{\alpha,z}}({\rho_0}:{\rho_{\tau}})}{2\,{\sum_l}\langle\langle{\Phi_{\alpha,z}}({\rho_0},{\rho_t}){\|{K_l}{\rho_0}(d{K_l^{\dagger}}/dt)\|_1} \rangle\rangle_{\tau}} ~,
\end{equation}
with ${\Phi_{\alpha,z}}({\rho_0},{\rho_t})$ defined in Eq.~\eqref{eq:0000000000012} accordingly.

We stress that Eq.~\eqref{eq:0000000000019} holds for entangled or separable, mixed quantum states. In particular, for $\alpha = z = 1/2$, Eq.~\eqref{eq:0000000000020} implies a QSL time written as a function of the Uhlmann fidelity. Next, for $z = 1$, it follows that ${\tau^{\text{NUQSL}}_{\alpha,z}}$ is recast in terms of the Petz-R\'{e}nyi relative entropy. We note that, by using the Pinsker inequality ${\text{R}_{\alpha}}({\rho_{\tau}}\|{\rho_0}) \geq (1/2){\|{\rho_{\tau}} - {\rho_0}\|_1^2}$, the latter result would imply a lower bound as a function of the trace distance between initial and final states of the dynamics, thus providing an extension of the results addressed in Ref.~\cite{bounding_generalized} to the case of nonunitary dynamics. For e\-xam\-ple, one finds that Eq.~\eqref{eq:0000000000020} is bounded from below as
\begin{align}
\label{eq:0000000000022}
&{\tau''_{\alpha,1}}({\rho_{\tau}}\|{\rho_0}) \geq \nonumber\\
& \frac{|1 - \alpha| \left|1 + (1 - \alpha)\ln[k_{\text{min}}({\rho_0})]\right| {\text{R}_{\alpha}}({\rho_{\tau}}\|{\rho_0})}{2 \alpha \, {\left[{k_{\text{max}}}({\rho_0})\right]^{1 - \alpha}} {\sum_l}\langle\langle {\left[{k_{\text{min}}}({\rho_t})\right]^{\alpha - 1}}{\|{K_l}{\rho_0}(d{K_l^{\dagger}}/dt)\|_1} \rangle\rangle_{\tau}} ~.
\end{align}

We note that Refs.~\cite{NewJPhys_24_065001_2022,NewJPhys_24_065003,arXiv:2303.07415,QSL_Entanglem003} investigate QSLs for open quantum systems by using relative entanglement entropy, Umegaki relative entropy, and quantum mutual information. These speed limits require evaluating the full spectrum of the Liouvillian super-operator that drives the nonunitary evolution of the quantum system. In turn, one finds that the entropic speed limit in Eq.~\eqref{eq:0000000000019} [see also Eqs.~\eqref{eq:0000000000020} and~\eqref{eq:0000000000021}] depends on the set of Kraus operators that generate the nonunitary dynamics, also being a function of the smallest and largest eigenvalues of states $\rho_0$ and $\rho_t$. The result in Eq.~\eqref{eq:0000000000019} may be useful, for example, in estimating the QSL time for open quantum systems whose dynamics is fully described by quantum channels, regardless of the strength of the interaction between system and environment.


\section{Examples}
\label{sec:00000000005}

In this section, we illustrate our findings consi\-de\-ring single-qubit and two-qubit states. In the former case, we investigate the unitary evolution generated by a time-independent Hamiltonian, also discussing a nonunitary evolution described by a depolarizing channel. In the latter, we address two-qubit states evolving under an amplitude damping channel.


\subsection{Single-qubit states}
\label{sec:00000000005A}

We consider a two-level system initialized at the single-qubit state ${\rho_0} = (1/2)(\mathbb{I} + \vec{r} \cdot \vec{\sigma})$, where $\vec{r} = r\left(\sin\theta\cos\phi,\sin\theta\sin\phi,\cos\theta\right)$ is the Bloch vector, with $r \in [0,1)$, $\theta \in [0,\pi]$, and $\phi\in[0,2\pi)$, while $\vec{\sigma} = \left({\sigma_x},{\sigma_y},{\sigma_z}\right)$ is the vector of Pauli matrices. We note that ${k_{\text{min}}}({\rho_0}) = (1 - r)/2$ and ${k_{\text{max}}}({\rho_0}) = (1 + r)/2$. In the following, we address both cases of unitary and nonunitary evolutions.


\subsubsection{Unitary evolution}
\label{sec:00000000005A1}

Let $H = \vec{n}\cdot\vec{\sigma}$ be the time-independent Hamiltonian of the two-level system, where $\vec{n} = ({n_x},{n_y},{n_z})$ is a three-dimensional vector. The unitary evolution is governed by the operator ${U_t} = {e^{- i t H}} = \cos({\|\vec{n}\|}t)\mathbb{I} - i\sin({\|\vec{n}\|}t)(\hat{n}\cdot\vec{\sigma})$, where ${\|\vec{n}\|^2} = {\sum_{l = x,y,z}}\,{n_l^2}$ is the Euclidean norm, with $\hat{n} = \vec{n}/\|\vec{n}\|$ and $\|\hat{n}\| = 1$. The instantaneous state is given by ${\rho_t} = {U_t}{\rho_0}{U^{\dagger}_t} = (1/2)[\mathbb{I} + r({\hat{r}_t}\cdot\vec{\sigma})]$, with ${\hat{r}_t} = \hat{r} + \sin(2{\|\vec{n}\|}t)(\hat{n}\times\hat{r}) + 2\,{\sin^2}({\|\vec{n}\|}t)[(\hat{n}\cdot\hat{r})\hat{n} - \hat{r}]$. The spectrum of the density matrix remains unchanged, i.e., ${k_{\text{min}/\text{max}}}({\rho_t}) = {k_{\text{min}/\text{max}}}({\rho_0})$, while the variance of the Hamiltonian $H$ with respect to the initial state ${\rho_0}$ is given by ${(\Delta{H})^2} = {\|\vec{n}\|^2}[1 - {(\hat{n}\cdot\vec{r}\,)^2}]$. In addition, the auxiliary function in Eq.~\eqref{eq:0000000000005} yields
\begin{equation}
\label{eq:0000000000024}
{h_{\alpha,z}}({\rho_0}) = \frac{{\left((1 + r){(1 - r)^{z - 1}} \right)^{\frac{1 - \alpha}{z}}}}{{2^{1 - \alpha}}\left|1 + (1 - \alpha)\ln\left(\frac{1 - r}{2}\right)\right|} ~.
\end{equation} 
In this setting, for $0 < \alpha < 1$, one finds that the $\alpha$-$z$-relative purity related to states $\rho_0$ and ${\rho_t}$ read
\begin{align}
\label{eq:0000000000022000000A}
&{g_{\alpha,z}}({\rho_t},{\rho_0}) = {\left({\epsilon_{\alpha,z}}(t)\right)^z}
\left[{\left(1 - \sqrt{1 - \frac{{(1 - {r^2})^{\frac{1}{z}}}}{{2^{\frac{2}{z}}}{\left({\epsilon_{\alpha,z}}(t)\right)^2}}} \,\right)^z} \right.\nonumber\\
&\left.+ {\left(1 + \sqrt{1 - \frac{{(1 - {r^2})^{\frac{1}{z}}}}{{2^{\frac{2}{z}}}{\left({\epsilon_{\alpha,z}}(t)\right)^2}}} \,\right)^z}\right] ~,
\end{align}
where we define 
\begin{equation}
\label{eq:0000000000022000000B}
{\epsilon_{\alpha,z}}(t) := \frac{1}{2}\left[{\xi_{\frac{1}{z}}^+} - {\xi_{\frac{\alpha}{z}}^-}{\xi_{\frac{1 - \alpha}{z}}^-}\left(1 - {(\hat{n}\cdot\hat{r})^2}\right){\sin^2}(\|\vec{n}\|t)\right] ~,
\end{equation}
with
\begin{equation}
\label{eq:0000000000022000000C}
{\xi_s^{\pm}} := {2^{-s}}\left[{(1 + r)^s} \pm {(1 - r)^s}\right] ~.
\end{equation}
In particular, for $t = {t_m} = m\pi/\|\vec{n}\|$, with $m \in \mathbb{N}$, we have that the auxiliary function in Eq.~\eqref{eq:0000000000022000000B} reduces to ${\epsilon_{\alpha,z}}({t_m}) = (1/2){\xi_{{1}/{z}}^+}$, which implies that ${g_{\alpha,z}}({\rho_{t_m}},{\rho_0}) = 1$, for all $\alpha \in (0,1)$, while the $\alpha$-$z$-RRE approaches zero.

The generalized entropic speed limit ${\tau^{\text{UQSL}}_{\alpha,z}}$ related to the unitary evolution of single-qubit states is obtained by combining Eqs.~\eqref{eq:0000000000014} [see also Eqs.~\eqref{eq:0000000000015} and~\eqref{eq:0000000000016}] and~\eqref{eq:0000000000022000000A} [see also Eqs.~\eqref{eq:0000000000022000000B} and~\eqref{eq:0000000000022000000C}]. It is worth noting that the QSL will be related to the quantum coherences of the probe state with respect to the fixed eigenbasis of the Hamiltonian. In fact, it turns out that Eq.~\eqref{eq:0000000000022000000B} depends on the quantity $1 - {(\hat{n}\cdot\hat{r})^2} = 2\,{\mathcal{I}_H}({\rho_0})/{({\|\vec{n}\|}r)^2}$, where ${\mathcal{I}_H}({\rho_0}) = -(1/4)\text{Tr}({[{\rho_0},H]^2})$ has been proven to be a useful coherence measure~\cite{PhysRevLett.113.170401}. Therefore, we have that $\rho_0$ is an incoherent state with respect to the eigenstates of $H$ for $\hat{n}\cdot\hat{r} = \pm 1$, while it exhibits non-zero quantum coherences in such a basis whenever $-1 < \hat{n}\cdot\hat{r} < 1$. The QSL trivially vanishes in the first case, but is expected to display nonzero values in the second.

Next, we investigate the QSL related to the Petz-R\'{e}nyi relative entropy (PRRE). In particular, fixing $z = 1$, we find that ${\xi^+_1} = 1$ and ${\xi^-_1} = r$ [see Eq.~\eqref{eq:0000000000022000000C}], which implies that ${\epsilon_{\alpha,1}}(t) = (1/2)[1 - {\xi_{\alpha}^-}{\xi_{1 - \alpha}^-}\left(1 - (\hat{n}\cdot\hat{r})^{2}\right){\sin^2}(\|\vec{n}\|t)]$. In this case, we have that the entropic QSL in Eq.~\eqref{eq:0000000000017} becomes
\begin{align}
\label{eq:0000000000022000000D}
&{\tau'_{\alpha,1}}({\rho_{\tau}}\|{\rho_0}) = \nonumber\\
& -\frac{ |1 + (1 - \alpha)\ln[(1 - r)/2]| \, \ln[{g_{\alpha,1}}({\rho_{\tau}},{\rho_0})]}{2 \, \alpha \, {(1 + r)^{1 - \alpha}}{(1 - r)^{\alpha - 1}}  {\|\vec{n}\|}\sqrt{1 - {(\hat{n}\cdot\vec{r}\,)^2}}} ~,
\end{align}
which is positive, real-valued, since $0 < {g_{\alpha,1}}({\rho_{\tau}},{\rho_0}) \leq 1$ [see Eq.~\eqref{eq:0000000000022000000A}]. It is important to note that Eq.~\eqref{eq:0000000000022000000D} partially recovers the QSL time ${\tau^R_{\alpha}}({\rho_{\tau}}\|{\rho_0})$ for PRRE that is reported in Ref.~\cite{bounding_generalized}; the latter result is valid for time-independent, unitary evolutions of single-qubit states. The ratio between these results is given by ${\tau'_{\alpha,1}}({\rho_{\tau}}\|{\rho_0})/{\tau^R_{\alpha}}({\rho_{\tau}}\|{\rho_0}) = {\Omega_{\alpha,r,\vartheta}}$, with
\begin{equation}
{\Omega_{\alpha,r,\vartheta}} = \frac{{\xi^-_{\alpha}}\sqrt{\xi^+_{2 - 2\alpha}}\,\sin\vartheta}{\alpha\sqrt{2(1 - {r^2}{\cos^2}\vartheta})}\,{\left(\frac{1 - r}{1 + r}\right)^{1 - \alpha}} ~,
\end{equation}
which in turn depends solely on parameters $\alpha$, $r$, and the angle $\vartheta \in [0,\pi]$ related to the unity vectors $\hat{n}$ and $\hat{r}$, i.e., $\hat{n}\cdot\hat{r} = \cos\vartheta$. We find that $0 \leq {\Omega_{\alpha,r,\vartheta}} \lesssim 1$, for all $\alpha \in (0,1)$, $r \in (0,1)$, and $\vartheta \in [0,\pi]$. This discrepancy comes from the fact that ${\tau^R_{\alpha}}({\rho_{\tau}}\|{\rho_0})$ in Ref.~\cite{bounding_generalized} was evaluated in terms of Schatten $2$-norm of the probe state, while here we find ${\tau'_{\alpha,1}}({\rho_{\tau}}\|{\rho_0})$ as a function of Schatten $1$-norm.


\subsubsection{Nonunitary evolution}
\label{sec:00000000005A2}

We consider the single-qubit state evolving under a depolarizing channel described by the time-dependent Kraus operators ${K_0} = (1/2)\sqrt{1 + 3\,{e^{-\Gamma t}}} \,\mathbb{I}$, and ${K_{1,2,3}} = (1/2)\sqrt{1 - {e^{-\Gamma t}}}\,{\sigma_{x,y,z}}$, with ${\sum_j}{K_j^{\dagger}}{K_j} = {\sum_j}{K_j}{K_j^{\dagger}} = \mathbb{I}$~\cite{NIELSEN}. Here, $\mathbb{I}$ is the $2\times2$ identity matrix, ${\sigma_j}$ is the $j$th Pauli matrix, and $\Gamma$ defines the characteristic dam\-ping rate of the Markovian nonunitary process. The ins\-tan\-ta\-neous state of the system is given by ${\rho_t} =  {\mathcal{E}_t}({\rho_0}) = {\sum_{j = 0}^3}{K_j}{\rho_0}{K_j^{\dagger}} = (1/2)\left(\mathbb{I} + {\vec{r}_t}\cdot\vec{\sigma}\right)$, with ${\vec{r}_t} = {e^{-\Gamma t}}\,\vec{r}$, whose eigenvalues are ${\kappa_{\text{min}}}({\rho_t}) = (1/2)(1 - {e^{-\Gamma t}}r)$ and ${\kappa_{\text{max}}}({\rho_t}) = (1/2)(1 + {e^{-\Gamma t}}r)$. From the geometry of single-qubit states, we have that the depolarizing channel homogeneously contracts the Bloch sphere toward the maximally mixed state located at its center.
\begin{figure}[t]
\begin{center}
\includegraphics[scale=1.1]{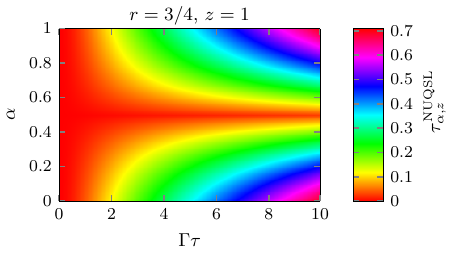}
\caption{(Color online) Density plot of the entropic QSL [see Eqs.~\eqref{eq:0000000000019},~\eqref{eq:0000000000020}, and~\eqref{eq:0000000000022}], as a function of the dimensionless parameters $\Gamma\tau$ and $\alpha$, for initial single-qubit states evolving under the depo\-la\-ri\-zing channel [see Sec.~\ref{sec:00000000005A}]. Here, we fix $z = 1$, and we set probe states with $r = 3/4$, for all $\theta \in [0,\pi]$ and $\phi \in [0,2\pi)$.}
\label{FIG000002}
\end{center}
\end{figure}

\begin{figure*}[!t]
\begin{center}
\includegraphics[scale=1]{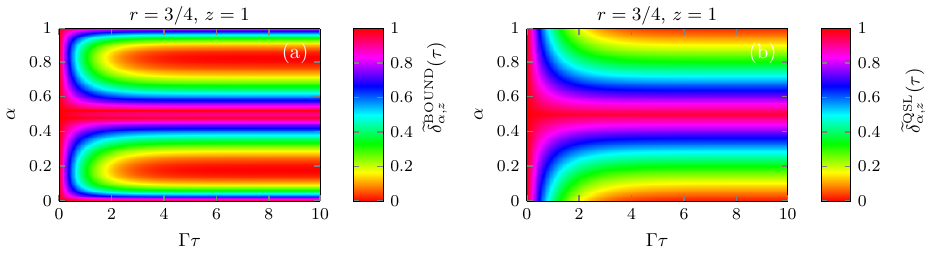}
\caption{(Color online) Density plot of the normalized relative errors ${\widetilde{\delta}^{\text{BOUND}}_{\alpha,z}}(\tau)$ [see Fig.~\ref{FIG03}(a)] and ${\widetilde{\delta}^{\text{QSL}}_{\alpha,z}}(\tau)$ [Fig.~\ref{FIG03}(b)], as a function of the dimensionless parameters $\Gamma\tau$ and $\alpha$, for a single-qubit state evolving under the depolarizing channel. Here we set $z = 1$, and we also choose the initial single-qubit state with $r = 3/4$.}
\label{FIG03}
\end{center}
\end{figure*}

In this setting, we find the analytical expression for $\alpha$-$z$-RRE given by
\begin{align}
\label{eq:0000000000023}
&{\text{D}_{\alpha,z}}({\rho_{\tau}}\|{\rho_0}) = \frac{1}{\alpha - 1}\left\{ \ln\left[ {(1 - r)^{\alpha - 1}}{(1 + {e^{-\Gamma \tau}}r)^{\alpha}} \right.\right.\nonumber\\
&\left.\left. + {(1 + r)^{\alpha - 1}}{(1 - {e^{-\Gamma \tau}}r)^{\alpha}} \right] - \ln\left[2{(1 - {r^2})^{\alpha - 1}}\right]\right\} ~.
\end{align}
Notably, Eq.~\eqref{eq:0000000000023} shows that the $\alpha$-$z$-RRE for single-qubit states subjected to the depolarizing process is independent of the parameters $z$, $\theta$, and $\phi$. In particular, the latter approaches the asymptotic value ${g_{\alpha,z}}({\rho_{\tau}},{\rho_0}) = (1/2)({(1 - r)^{1 - \alpha}} + {(1 + r)^{1 - \alpha}})$ in the limiting case $\Gamma\tau \rightarrow \infty$. This implies that the $\alpha$-$z$-RRE will be fully characterized by the parameters $\alpha$ and $r$ at later times of the dynamics. In addition, we find the upper bound on the Schatten speed as ${\|{K_j}{\rho_0}(d{K_j^{\dagger}}/d{t})\|_1} = (\Gamma/8)(1 + 2{\delta_{j0}}){e^{-\Gamma t}}$, for all $j = \{0,1,2,3\}$. Therefore, the entropic QSL ${\tau^{\text{NUQSL}}_{\alpha,z}}$ in Eq.~\eqref{eq:0000000000019} is evaluated by combining these results with Eqs.~\eqref{eq:0000000000024} and~\eqref{eq:0000000000023}. In detail, from Eq.~\eqref{eq:0000000000020}, we get the result
\begin{align}
\label{eq:0000000000025}
&{\tau''_{\alpha,z}}({\rho_{\tau}}\|{\rho_0}) = 2 \tau r \left|1 + (1 - \alpha)\ln\left[{(1 - r)}/{2}\right]\right|\times\nonumber\\
&\times \frac{\ln\left[\frac{2 \, {(1 - {r^2})^{\alpha - 1}}}{{(1 - r)^{\alpha - 1}}{(1 + {e^{-\Gamma \tau}}r)^{\alpha}} + {(1 + r)^{\alpha - 1}}{(1 - {e^{-\Gamma \tau}}r)^{\alpha}}}\right]}{3 \, {\left[{(1 - r)^{z - 1}}(1 + r)\right]^{\frac{1 - \alpha}{z}}} \left[{(1 - {e^{-\Gamma\tau}}r)^{\alpha}} - {(1 - r)^{\alpha}}\right]} ~,
\end{align}
with ${\tau''_{\alpha,z}}({\rho_{\tau}}\|{\rho_0}) = {\tau''_{1 - \alpha,z}}({\rho_0}\|{\rho_{\tau}})$. We also obtain ana\-ly\-tical result for ${\tau''_{\alpha,z}}({\rho_{\tau}}:{\rho_0})$ defined in Eq.~\eqref{eq:0000000000021}, but it is too cumbersome to be reported here. Overall, we verify that the entropic QSL is an increasing function for $1 \geq z \geq \text{max}\{\alpha, 1 - \alpha\}$, i.e., ${\tau^{\text{NUQSL}}_{\alpha,{z_1}}} > {\tau^{\text{NUQSL}}_{\alpha,{z_2}}}$ for ${z_1} > {z_2}$.

In Fig.~\ref{FIG000002}, we plot the entropic QSL time in Eq.~\eqref{eq:0000000000019} [see Eqs.~\eqref{eq:0000000000020},~\eqref{eq:0000000000021},~\eqref{eq:0000000000023}, and~\eqref{eq:0000000000024}], as a function of the dimensionless parameters $\alpha$ and $\Gamma\tau$, for single-qubit states subjected to the depolarizing channel. Here we set $z = 1$, and choose the initial state with mixing parameter $r = 3/4$, for all $\theta \in [0,\pi]$ and $\phi \in [0,2\pi)$. We see that the QSL time exhibits the symmetry ${\tau^{\text{NUQSL}}_{\alpha,z}} = {\tau^{\text{NUQSL}}_{1 - \alpha,z}}$, for all $\Gamma\tau \geq 0$. Note that, for a fixed value $\alpha\in (0,1/2)\cup(1/2,1)$, the entropic speed limit increases as a function of the dimensionless parameter $\Gamma\tau$. In particular, for $\alpha = 1/2$, we see that the QSL time exhibits small values for all $\Gamma\tau \geq 0$. In this case, one verifies that the $\alpha$-$z$-relative entropy approaches zero, i.e., ${\text{D}_{1/2,z}}({\rho_{\tau}}\|{\rho_0}) \approx 0$ for all $\tau \geq 0$, since the relative purity is close to unity. Hence, we have that ${\tau^{\text{NUQSL}}_{1/2,z}} \approx 0$ along the noisy dynamics.

Furthermore, Fig.~\ref{FIG000002} shows that ${\tau^{\text{NUQSL}}_{\alpha,z}} \ll 1$ for $0 \leq \Gamma\tau \lesssim 2$ and $\alpha \in (0,1)$. In fact, for $\Gamma\tau \ll 1$, we verify that ${\text{D}_{\alpha,z}}({\rho_{\tau}}\|{\rho_0}) \approx {\alpha}{r^2}{(\Gamma\tau)^2}/[2(1 - {r^2})]$ scales quadratically with $\Gamma\tau$, which implies that the QSL approaches small values at earlier times of the dynamics. In turn, for a fixed value of $\Gamma\tau \gtrsim 2$, one finds that the QSL time is non-monotonic as a function of the parameter $\alpha \in (0,1)$. In this regime, it decreases for $0 \lesssim \alpha < 1/2$, fluctuates around zero for $\alpha = 1/2$, and increases for $1/2 < \alpha < 1$. This behavior is more pronounced at later times of the dynamics.

In the following, we discuss the relative errors in Eqs.~\eqref{eq:0000000000008001} and~\eqref{eq:000000000001302} for the single-qubit state evolving nonunitarily under the depolarizing channel. For our purposes, throughout the paper we will focus on the normalized relative error ${\widetilde{\delta}_{\alpha,\mu}}(\tau)$, with $\widetilde{x} := (x - \min(x))/(\max(x) - \min(x))$, noting that $0 \leq {\widetilde{\delta}_{\alpha,\mu}}(\tau) \leq 1$. In Figs.~\ref{FIG03}(a) and~\ref{FIG03}(b), we plot the normalized relative errors ${\widetilde{\delta}^{\text{BOUND}}_{\alpha,z}}$ and ${\widetilde{\delta}^{\text{QSL}}_{\alpha,z}}$, respectively, as a function of the dimensionless parameters $\Gamma\tau$ and $\alpha$. The density plots are symmetric with respect to the mapping $\alpha \rightarrow 1 - \alpha$, inherited from the symmetrized $\alpha$-$z$-RRE.

On the one hand, Fig.~\ref{FIG03}(a) shows that $0 \lesssim {\widetilde{\delta}^{\text{BOUND}}_{\alpha,z}} \lesssim 0.2$ for $0.1 \lesssim \alpha \lesssim 0.3$ (also for $0.7 \lesssim \alpha \lesssim 0.9$), with $\Gamma\tau \gtrsim 2$. In turn, for $0.45 \lesssim \alpha \lesssim 0.55$, the bound is looser regardless of $\Gamma\tau \geq 0$. In particular, we find that ${\widetilde{\delta}^{\text{BOUND}}_{\alpha,z}} \approx 1$ for $0 \leq \Gamma\tau \lesssim 0.15$, for all $\alpha\in(0,1)$, thus implying that the bound is looser at earlier times of the nonunitary dynamics. On the other hand, for $\Gamma\tau \gtrsim 0.25$, Fig.~\ref{FIG03}(b) shows that the relative error ${\widetilde{\delta}^{\text{QSL}}_{\alpha,z}}$ increases (decreases) over the interval $0 < \alpha \lesssim 0.45$ ($0.55 \lesssim \alpha < 1$). In particular, one finds that $0 \lesssim {\widetilde{\delta}^{\text{QSL}}_{\alpha,z}} \lesssim 0.2$ for $0 < \alpha \lesssim 0.1$ (also for $0.9 \lesssim \alpha \lesssim 1$), with $\Gamma\tau \gtrsim 2$. We note that, for $0 \leq \Gamma\tau \lesssim 0.25$, one obtains the trivial bound ${\widetilde{\delta}^{\text{QSL}}_{\alpha,z}} \gtrsim 0$ for all $\alpha \in (0,1)$. Furthermore, the QSL bound is looser for $0.45 \lesssim \alpha \lesssim 0.55$, for all $\Gamma\tau \geq 0$.

\begin{figure*}[!t]
\begin{center}
\includegraphics[scale=1]{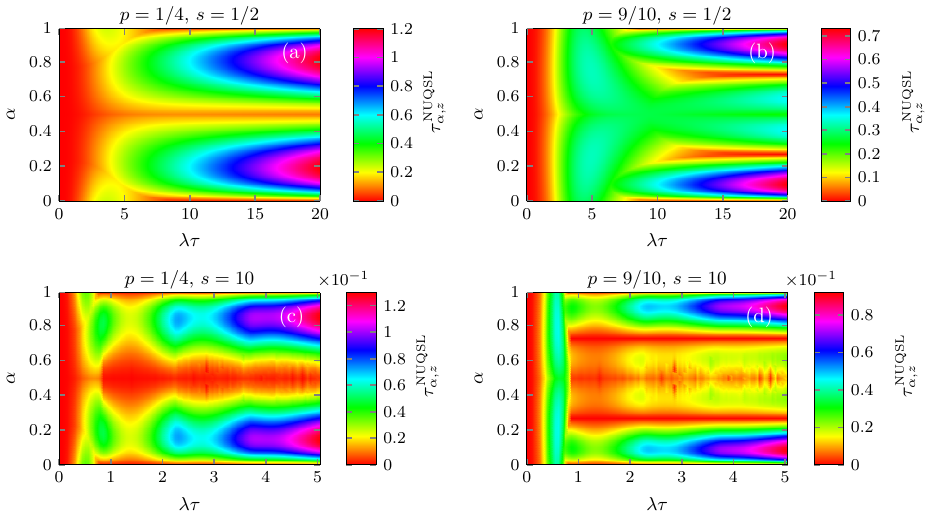}
\caption{(Color online) Density plot of the entropic QSL time in Eq.~\eqref{eq:0000000000019} [see Eqs.~\eqref{eq:0000000000020}, and~\eqref{eq:0000000000021}], as a function of the dimensionless parameters $\alpha$ and $\lambda\tau$, for the system of two-qubits coupled to independent and identical environments [see Sec.~\ref{sec:00000000005B}]. We choose $z = 1$, and we also consider the mixing parameters with respect to separable states, $p = 1/4$ [see Figs.~\ref{FIG000004}(a) and~\ref{FIG000004}(c)], and entangled initial GHZ states, $p = 9/10$ [see Figs.~\ref{FIG000004}(b) and~\ref{FIG000004}(d)]. In addition, we set the effective amplitude damping dynamics ranging from the Markovian regime, $s = 1/2$ [see Figs.~\ref{FIG000004}(a) and~\ref{FIG000004}(b)], to the non-Markovian regime, $s = 10$ [see Figs.~\ref{FIG000004}(c) and~\ref{FIG000004}(d)].}
\label{FIG000004}
\end{center}
\end{figure*}


\subsection{Two-qubit states}
\label{sec:00000000005B}

Next, we consider a system of two qubits in\-terac\-ting with identical, noncorrelated, independent reservoirs at zero temperature~\cite{Amplit_Damping_Palma}. The system is initialized at the mixed state ${\rho_0} = [(1 - p)/4] (\mathbb{I}\otimes\mathbb{I}) + p|{\text{GHZ}_2}\rangle\langle{\text{GHZ}_2}|$, with $0 \leq p \leq 1$, where $|{\text{GHZ}_2}\rangle := (1/\sqrt{2})\left(|0,0\rangle + |1,1\rangle\right)$ is the Greenberger-Horne-Zeilinger (GHZ) state. Here $\{|0\rangle,|1\rangle\}$ define the computational basis states in the complex two-dimensional vector space ${\mathbb{C}^2}$. The noisy processes of energy dissipation and decoherence are modeled by the amplitude-damping channel~\cite{Amplit_damping002}. The instantaneous state of the two-qubits system is given by ${\rho_t} = {\mathcal{E}_t}({\rho_0}) = {\sum_{j,l = 1,2}}\left({K_j}\otimes{K_l}\right){\rho_0}({K^{\dagger}_j}\otimes{K^{\dagger}_l})$, with the time-dependent Kraus operators ${K_1} = |0\rangle\langle0| + {\gamma_t}|1\rangle\langle{1}|$, and ${K_2} = \sqrt{1 - {|{\gamma_t}|^2}} \, |0\rangle\langle{1}|$, while ${\sum_{j = 1,2}}{K^{\dagger}_j}{K_j} = \mathbb{I}$. In the resonant regime, that is, when the frequency of the qubits coincides with the peak frequency of the reservoirs, it is known that the decoherence function $\gamma_t$ is given by~\cite{Breuer_Petruccione}
\begin{equation}
\label{eq:0000000000026}
{\gamma_t} = {e^{-\lambda t/2}}\left[\cosh\left(\frac{\lambda t}{2}\sqrt{1 - 2s}\right) + \frac{\sinh\left(\frac{\lambda t}{2}\sqrt{1 - 2s} \,\right)}{\sqrt{1 - 2s}}\right] ~,
\end{equation}
where $\lambda$ is the width of the spectral density of the reservoirs, while the $s \geq 0$ is a dimensionless parameter that sets Markovian ($0 \leq s \leq 1/2$), and non-Markovian ($s > 1/2$) regimes of the dynamical map~\cite{PhysRevA.84.031602}. It can be shown that, for all $t \geq 0$ and $s \geq 0$, the absolute value of Eq.~\eqref{eq:0000000000026} is bounded as $0\leq |{\gamma_t}|\leq 1$, and also ${\gamma_t} = {e^{-\lambda t/2}}(1 + \lambda t/2)$ remains finite for $s \rightarrow 1/2$. Furthermore, one finds that ${\gamma_t}\approx 0$ for $\lambda t \gg 1$, which implies that the instantaneous state ${\rho_t}\approx |0\rangle\langle0|\otimes|0\rangle\langle0|$ becomes asymptotically separable at later times of the dynamics.

\begin{figure*}[!t]
\begin{center}
\includegraphics[scale=1]{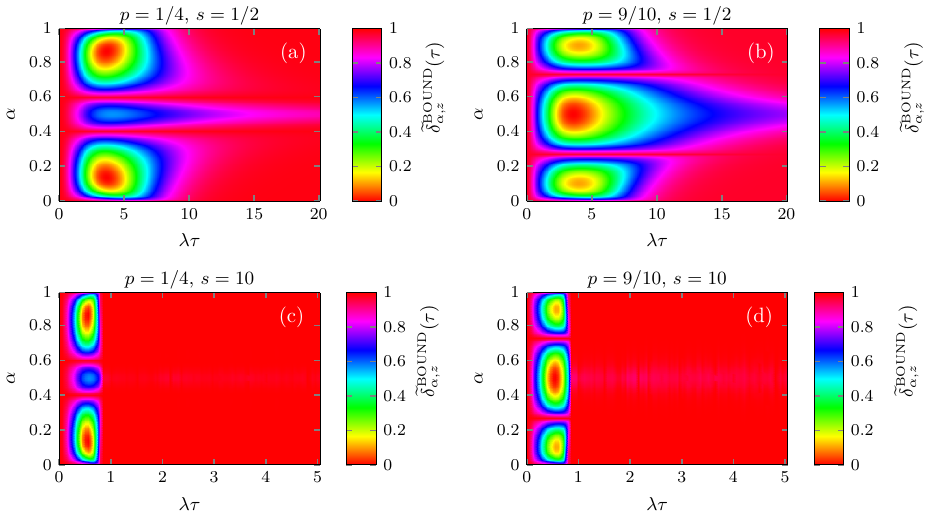}
\caption{(Color online) Density plot of the normalized relative error ${\widetilde{\delta}^{\text{BOUND}}_{\alpha,z}}(\tau)$, as a function of the dimensionless parameters $\lambda\tau$ and $\alpha$, for the system of two-qubits evolving under the amplitude damping channel. Here we set $z = 1$, and we also consider the mixing parameters with respect to separable states, $p = 1/4$ [see Figs.~\ref{FIG000005}(a) and~\ref{FIG000005}(c)], and entangled initial GHZ states, $p = 9/10$ [see Figs.~\ref{FIG000005}(b) and~\ref{FIG000005}(d)]. In addition, we set the effective amplitude damping dynamics ranging from the Markovian regime, $s = 1/2$ [see Figs.~\ref{FIG000005}(a) and~\ref{FIG000005}(b)], to the non-Markovian regime, $s = 10$ [see Figs.~\ref{FIG000005}(c) and~\ref{FIG000005}(d)].}
\label{FIG000005}
\end{center}
\end{figure*}

The spectrum of the initial state is given by ${k_{\text{max}}}({\rho_0}) = (1 + 3p)/4$, and also the three-fold degenerate eigenvalue ${k_{\text{min}}}({\rho_0}) = (1 - p)/4$. We point out that the probe state is separable whenever $p \leq 1/3$. In turn, the instantaneous state exhibits the smallest eigenvalue 
\begin{align}
\label{eq:0000000000027}
&{k_{\text{min}}}({\rho_t}) = \frac{1}{2}\left(1 - \sqrt{1 - {\gamma_t^2}(2 - (1 + {p^2}){\gamma_t^2})}\right. \nonumber\\
&\left. - \frac{1}{2}{\gamma_t^2}(2 - (1 + p){\gamma_t^2})\right) ~,
\end{align}
which in turn is threefold-degenerate at earlier and later times of the dynamics. In addition, after straightforward calculations, one finds the Schatten $1$-norm involving the rate of change of Kraus operators as follows
\begin{align}
\label{eq:0000000000028}
&{\sum_{j,l = 1,2}}{\|{K_{jl}}{\rho_0}(d{K_{jl}^{\dagger}}/dt)\|_1} = \frac{1}{2}\left|{\gamma_t}\frac{d{\gamma_t}}{dt}\right|\left[2(1 - p) \right.\nonumber\\
&\left.+ (1 + p)\left(\left|1 - {\gamma_t^2}\right|+\left|1 - 2{\gamma_t^2}\right|\right) + \sqrt{4{p^2} + {(1 + p)^2}{\gamma_t^4}} \, \right] ~,
\end{align}
with ${K_{jl}} := {K_j}\otimes{K_l}$, and ${d}{\gamma_t}/d{t} = -\left(\lambda s/\sqrt{1 - 2s} \, \right)e^{-\lambda t/2}\sinh\left(\frac{\lambda t}{2}\sqrt{1 - 2s} \, \right)$. The analytical calculation of $\alpha$-$z$-RRE involves expressions too long to include here. In the following, we provide numerical simulations of the entropic speed limit.

Figure~\ref{FIG000004} shows the phase diagram of the QSL time in Eq.~\eqref{eq:0000000000019} [see Eqs.~\eqref{eq:0000000000020}, and~\eqref{eq:0000000000021}], as a function of the dimensionless parameters $\alpha$ and $\lambda\tau$, for the system of two-qubits coupled to independent identical reservoirs. We set $z = 1$, and we consider the mixing parameters with respect to separable states, $p = 1/4$ [see Figs.~\ref{FIG000004}(a) and~\ref{FIG000004}(c)], and entangled initial states, $p = 9/10$ [see Figs.~\ref{FIG000004}(b) and~\ref{FIG000004}(d)]. In addition, we set the effective amplitude damping dynamics ranging from the Markovian regime, $s = 1/2$ [see Figs.~\ref{FIG000004}(a) and~\ref{FIG000004}(b)], to the non-Markovian regime, $s = 10$ [see Figs.~\ref{FIG000004}(c) and~\ref{FIG000004}(d)]. The QSL time exhibits decreasing intensities from memoryless ($s = 1/2$) to memory-bearing ($s = 10$) regimes of the nonunitary evolution, somehow speeding up the dynamics. Overall, one verifies the symmetry ${\tau^{\text{NUQSL}}_{\alpha,z}} = {\tau^{\text{NUQSL}}_{1 - \alpha,z}}$, for all $\lambda \tau \geq 0$. In addition, for all $\alpha \in (0,1)$, the QSL time approaches small values at earlier times of the dynamics.

In Figs.~\ref{FIG000004}(a),~\ref{FIG000004}(c), and~\ref{FIG000004}(d), we see that the entropic speed limit approaches small values around the critical line $\alpha = 1/2$, for all $\lambda\tau \geq 0$. In turn, Fig.~\ref{FIG000004}(b) shows that the QSL takes non-vanishing values $0.2 \lesssim {\tau^{\text{NUQSL}}_{\alpha,z}} \lesssim 0.3$, for $0.4 \lesssim \alpha \lesssim 0.6$. We note that, fixing the value of $0 < \alpha < 1$ ($\lambda\tau \geq 0$), the QSL time is non-monotonic as a function of the parameter $\lambda\tau \geq 0$ ($0 < \alpha < 1$). This is a typical fingerprint of non-Markovian dynamics, and we see that the entropic QSL is able to capture such a signature. The transition from Markovian [see Figs.~\ref{FIG000004}(a) and~\ref{FIG000004}(b)] to non-Markovian dynamics [see Figs.~\ref{FIG000004}(c) and~\ref{FIG000004}(d)] contracts the non-vanishing region of QSL times in the phase diagram. In addition, we see that the entropic speed limit exhibits nontrivial fluctuations in the non-Markovian regime [see Figs.~\ref{FIG000004}(c) and~\ref{FIG000004}(d)]. This behavior is inherited from the decoherence function, which in turn becomes an oscillating function for $s > 1/2$.

\begin{figure*}[!t]
\begin{center}
\includegraphics[scale=1]{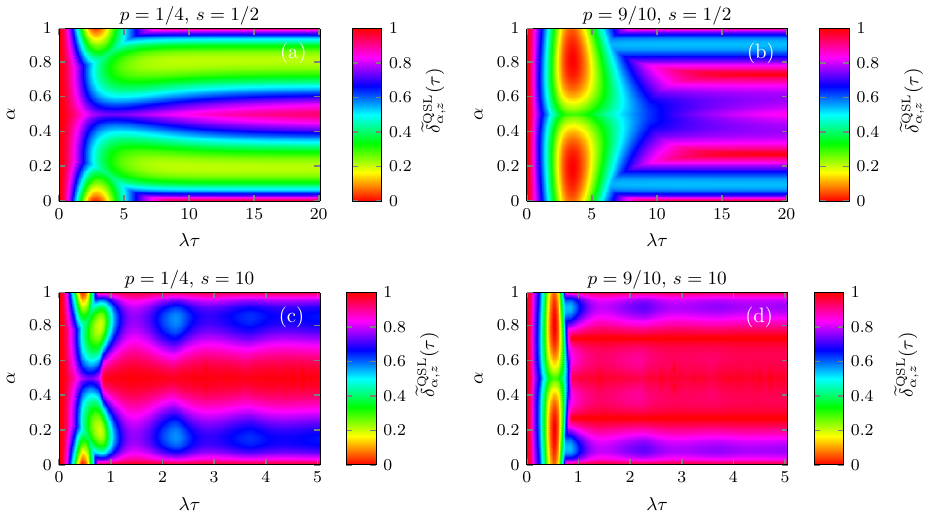}
\caption{(Color online) Density plot of the normalized relative error ${\widetilde{\delta}^{\text{QSL}}_{\alpha,z}}(\tau)$, as a function of the dimensionless parameters $\lambda\tau$ and $\alpha$, for the system of two-qubits evolving under the amplitude damping channel. Here we set $z = 1$, and we also consider the mixing parameters with respect to separable states, $p = 1/4$ [see Figs.~\ref{FIG000006}(a) and~\ref{FIG000006}(c)], and entangled initial GHZ states, $p = 9/10$ [see Figs.~\ref{FIG000006}(b) and~\ref{FIG000006}(d)]. In addition, we set the effective amplitude damping dynamics ranging from the Markovian regime, $s = 1/2$ [see Figs.~\ref{FIG000006}(a) and~\ref{FIG000006}(b)], to the non-Markovian regime, $s = 10$ [see Figs.~\ref{FIG000006}(c) and~\ref{FIG000006}(d)]}
\label{FIG000006}
\end{center}
\end{figure*}

In Figs.~\ref{FIG000005} and~\ref{FIG000006}, we show the density plots of the normalized relative errors ${\widetilde{\delta}^{\text{BOUND}}_{\alpha,z}}$ and ${\widetilde{\delta}^{\text{QSL}}_{\alpha,z}}$, respectively, as a function of the dimensionless parameters $\alpha$ and $\lambda\tau$, for the system of two-qubits coupled to independent identical reservoirs. We set $z = 1$, and we consider the mixing parameters with respect to separable states, $p = 1/4$ [see Figs.~\ref{FIG000005}(a) and~\ref{FIG000005}(c), Figs.~\ref{FIG000006}(a) and~\ref{FIG000006}(c)], and entangled initial states, $p = 9/10$ [see Figs.~\ref{FIG000005}(b) and~\ref{FIG000005}(d), Figs.~\ref{FIG000006}(b) and~\ref{FIG000006}(d)]. We also investigate the cases of Markovian nonunitary dynamics, $s = 1/2$ [see Figs.~\ref{FIG000005}(a) and~\ref{FIG000005}(b), Figs.~\ref{FIG000006}(a) and~\ref{FIG000006}(b)], and non-Markovian nonunitary dynamics, $s = 10$ [see Figs.~\ref{FIG000005}(c) and~\ref{FIG000005}(d), Figs.~\ref{FIG000006}(c) and~\ref{FIG000006}(d)].

On the one hand, for the case of Markovian dynamics, $s = 1/2$, Figs.~\ref{FIG000005}(a) and~\ref{FIG000005}(b) show that the bound is looser for separable initial mixed states ($p = 1/4$) than for entangled mixed GHZ states ($p = 9/10$). In Fig.~\ref{FIG000005}(a), one finds that $0.6 \lesssim {\widetilde{\delta}^{\text{BOUND}}_{\alpha,z}} \lesssim 0.8$ for $1 \lesssim \lambda\tau \lesssim 20$ and $0.45 \lesssim \alpha \lesssim 0.55$. In turn, Fig.~\ref{FIG000005}(b) shows that, for $1 \lesssim \lambda\tau\lesssim 20$, one finds $0 \lesssim {\widetilde{\delta}^{\text{BOUND}}_{\alpha,z}} \lesssim 0.8$ for $0.35 \lesssim \alpha \lesssim 0.75$. On the other hand, for the case of non-Markovian dynamics, we find that the bound is looser at later times of the nonunitary evolution. We note that, for $s > 1/2$ and $\tau \gg {\lambda^{-1}}$, the evolved state of the two-qubits system approaches the separable state $|{0}\rangle\langle{0}|\otimes|{0}\rangle\langle{0}|$, which in turn has a vanishing zero-valued smallest eigenvalue, i.e., ${\kappa_{\text{min}}}({\rho_{\tau}}) \approx {\kappa_{\text{min}}}(|{0}\rangle\langle{0}|\otimes|{0}\rangle\langle{0}|) \approx 0$. Hence, for $\lambda\tau \gtrsim 1$ and $\alpha \in (0,1)$, the upper bound on $\alpha$-$z$-RRE [see Eq.~\eqref{eq:0000000000007}] will assume larger values, thus implying a looser bound.

In Fig.~\ref{FIG000006}(a), one verifies that $0.2 \lesssim {\widetilde{\delta}^{\text{QSL}}_{\alpha,z}} \lesssim 0.8$ for $\lambda\tau \geq 1$ and $0 < \alpha \lesssim 0.4$ (also $0.6 \lesssim \alpha < 1$), for initial se\-pa\-ra\-ble mixed GHZ states ($p = 1/4$) evolving under a Markovian amplitude damping channel. In Fig.~\ref{FIG000006}(b), we have that the QSL bound is tighter at earlier times of the nonunitary dynamics, with $0 \lesssim {\widetilde{\delta}^{\text{QSL}}_{\alpha,z}} \lesssim 0.8$ for $1 \lesssim \lambda\tau \lesssim 6$, and $0 < \alpha < 1$. Figures~\ref{FIG000006}(c) and~\ref{FIG000006}(d) show that the entropic QSL bound is tighter at earlier times for initial mixed GHZ states evolving under a non-Markovian amplitude damping channel. In fact, we have that $0 \lesssim {\widetilde{\delta}^{\text{QSL}}_{\alpha,z}} \lesssim 0.4$ for $0.25 \lesssim \lambda\tau \lesssim 0.75$, with $\alpha \in (0,1)$ [see Fig.~\ref{FIG000006}(d)]. In both panels, one finds the normalized relative error ${\widetilde{\delta}^{\text{QSL}}_{\alpha,z}} \approx 1$ for $\lambda\tau \gg 1$, which means that the QSL time recovers the trivial bound $\tau \geq 0$ at later times of the dynamics, regardless of $\alpha \in (0,1)$.
 

\section{Conclusions}
\label{sec:00000000006}

In this work, we have discussed generalized entropic quantum speed limits based on $\alpha$-$z$-R\'{e}nyi relative entropies for finite-dimensional quantum systems evolving under arbitrary physical dynamics. The main contribution relies on the derivation of a class of QSLs valid for both closed and open quantum systems, i.e., unitary and nonunitary evolutions. To do so, we investigate the rate of change of $\alpha$-$z$-RRE evaluated for the probe and instantaneous states of the system, thus providing a family of two-parameter upper bounds on this entropic measure. We consider both non-symmetric and symmetric versions of $\alpha$-$z$-RREs. The bounds depend on the smallest and largest eigenvalues of these density matrices, as well as the Schatten speed of the instantaneous state. The latter quantity captures the very notion of the speed related to the dynamical map that governs the evolution of the quantum system. We note that the bounds require low computational cost, also encompassing several particular results depending on the choice of parameters $\alpha \in (0,1)$ and $1 \geq z \geq \text{max}\{\alpha, 1 - \alpha\}$.

We investigate the link between the quantum speed limit and $\alpha$-$z$-RREs. The bounds apply for general physical processes, mixed and pure states, and entangled or separable states. On the one hand, the family of entropic QSLs depend on such a family of generalized entropies, i.e., the distinguishability between probe and target states of the dynamics. Indeed, the farther apart (closer) these two states are, the greater (smaller) the entropic QSL will be. On the other hand, the QSL time depends on the time-average of the Schatten speed of the instantaneous state, weighted by a function of its smallest eigenvalue. We provide a unified entropic QSL by taking the maximum of the speed limits obtained from non-symmetric and symmetric  $\alpha$-$z$-RREs. Importantly, the entropic QSL exhibits the symmetry ${\tau_{\alpha,z}^{\text{QSL}}} = {\tau_{1 - \alpha,z}^{\text{QSL}}}$ with respect to the mapping $\alpha \rightarrow 1 - \alpha$, which in turn is inherited from the generalized family of entropies.

We have specialized these results to the case of unitary and nonunitary evolutions, and we obtained two classes of speed limits accordingly. On the one hand, for closed systems, the QSL time scales with the inverse of the variance of the time-independent Hamiltonian of the system, also being a function of both smallest and largest eigenvalues of the initial state. We find that the QSL is sensitive to the distinguishability between probe and target states, and it captures the fluctuations of the generator of the dynamics. In particular, for $z = 1$, we find the speed limit partially recovers the family of QSLs addressed in Ref.~\cite{bounding_generalized} for the case of time-independent Hamiltonians. On the other hand, we consider open quantum systems whose effective dynamics is given by CPTP maps. We note that the QSL depends on the Schatten $1$-norm related to the rate of change of the Kraus operators with respect to the quantum channel. In particular, for $z = 1$, the QSLs depend on Petz-R\'{e}nyi relative entropy and somehow promote an extension of the results of Ref.~\cite{bounding_generalized} to the case of nonunitary dynamics.

We illustrate our results for the case of single-qubit and two-qubit states. On the one hand, for single-qubit states, we set a two-level system undergoing unitary and nonunitary evolutions. The unitary evolution is governed by a time-independent Hamiltonian, and the QSL time depends on the coherences of an initial state with respect to the eigenbasis of such an operator. For the nonunitary evolution, we set a depolarizing channel. We see that, for a fixed $\alpha \in (0,1)$, the entropic QSL is monotonic as a function of time. In addition, it exhibits small values at earlier times of the dynamics, while it fluctuates around zero over the critical line $\alpha = 1/2$. On the other hand, for two-qubit states, we discuss the entropic QSL for a system of two-qubits coupled to independent reservoirs in the resonant regime. We consider an initial mixed GHZ state, analyzing both separable and entangled settings. We provide numerical simulations to the QSL, showing a phase diagram that depends on the mixing degree of the initial state, and also on the dimensionless parameter that controls both Markovian and non-Markovian regimes of the dynamics. We find that the QSL exhibits a rich phase diagram, thus being non-monotonic as a function of the time.

Our analysis revealed that the bounds are not tight for the dynamics considered. This may be related to the inequalities we used to bound $\alpha$-$z$-RRE above. In fact, tighter bounds are expected by applying a small number of inequalities. However, this can lead to intricate results, the evaluation of which may require a lot of prior information, especially when considering higher-dimensional systems. We emphasize that our bounds require minimal information about the physical system, e.g., the smallest and largest eigenvalues of the initial and evolved states, and also details about the generator of the dynamics. Our bounds are able to provide an estimate of $\alpha$-$z$-RRE since they require low computational cost, although they are not the most accurate.

The bounds can be experimentally investigated, for example, by exploring nuclear spins on Nuclear Magnetic Resonance (NMR) platforms with an ensemble of chloroform molecules ($\text{CHCl}_3$) in liquid-state at room temperature. We mention the experimental study of geometric QSLs discussed in Ref.~\cite{QSL_Experimental002}. The speed limit is related to the nonunitary evolution of a two-level system imprinted on carbon nuclear spins, where hydrogen nuclear spins become a decoherence source for the carbon nuclear spins. These spins, initially polarized along the direction of a static magnetic field, precess as they are disturbed by a weak oscillating magnetic field, also called a radio frequency pulse. This mechanism is related to the return to equilibrium, which in turn induces a voltage in a superconducting coil, captured in terms of the total magnetization of the nuclear spins.

The relaxation processes of the longitudinal $\langle{\sigma_z}\rangle_t = \text{Tr}({\sigma_z}{\rho_t})$ and transverse $\langle{\sigma_{x,y}}\rangle_t = \text{Tr}({\sigma_{x,y}}{\rho_t})$ magnetizations occur on time scales $T_1$ and $T_2$, respectively, where $\rho_t$ is the instantaneous state of the qubit encoded in the carbon nuclear spins. The former is related to the populations, while the latter is related to the coherences of the quantum state. The experimental procedure requires measurements of the spin magnetizations $\langle{\sigma_{x,y,z}}\rangle_t$ of single-qubit states by means of quantum tomography. To compare the experimental data with the theoretical results for single-qubit states, one can recast the entropic QSL bounds in terms of the spin magnetizations $\langle{\sigma_{x,y,z}}\rangle_t$. We consider the states ${\rho_{0,t}} = (1/2)(\mathbb{I} + {\langle\vec{\sigma}\rangle_{0,t}}\cdot\vec{\sigma})$, with ${\langle\vec{\sigma}\rangle_{0,t}} = ({\langle{\sigma_x}\rangle_{0,t}},{\langle{\sigma_y}\rangle_{0,t}},{\langle{\sigma_z}\rangle_{0,t}})$. Next, we rewrite the bounds ${\tau^{\text{NUQSL}}_{\alpha,z}}$ [see Eq.~\eqref{eq:0000000000019}], ${\tau''_{\alpha,z}}({\rho_{\tau}}\|{\rho_0})$ [see Eq.~\eqref{eq:0000000000020}], ${\tau''_{\alpha,z}}({\rho_0}:{\rho_{\tau}})$ [see Eq.~\eqref{eq:0000000000021}] in terms of the single-qubit magnetizations $\langle{\sigma_{x,y,z}}\rangle_{0,t}$. In this case, it is possible to investigate the QSL bounds as a function of the magnetizations accessible from the experimental data. These results are then compared with the numerical simulation of the bounds.

We emphasize that our results apply for quantum systems undergoing an arbitrary physical process, thus providing a two-parameter family of generalized entropic quantum speed limits. This general QSL encodes distinct classes of bounds on the evolution time, which in turn are decoded by specifying the $\alpha$-$z$-RRE, and also by discriminating the overall dynamics of the system. Interestingly, our results may recover previous QSLs reported in the lite\-ra\-ture using suitable inequalities that bound $\alpha$-$z$-RRE from above or below. Still undetermined is the link between our results and the so-called unified $(\alpha,z)$-relative entropy discussed in Ref.~\cite{IntJTheorPhys_50_1282}. Furthermore, one can investigate the entropic QSLs to other regions over the parameter space $(\alpha,z)$ where the entropy satisfies the data processing inequality, namely, $\alpha \in (0,1)$ and $z \geq 1$, $\alpha \in (1,2)$ and $z \in [\alpha/2,\alpha]$, $\alpha \in [2,\infty)$ and $z \in [\alpha - 1,\alpha]$. We hope to address these points in further investigations. Our results may find applications in the study of entropy production in the relaxation process at finite temperature, and they may also be useful in the study of the dynamics of quantum resources such as coherence and entanglement~\cite{IEEE_65_5880_2019}. Finally, our results may contribute to the study of thermodynamic and energy-time uncertainty relations~\cite{arXiv:2404.18163,Renyi_Uncert_Relation}.


\begin{acknowledgments}
This work was supported by the Brazilian ministries MEC and MCTIC, and the Brazilian funding agencies CNPq, and Coordena\c{c}\~{a}o de Aperfei\c{c}oamento de Pessoal de N\'{i}vel Superior--Brasil (CAPES) (Finance Code 001). J. F. S. and D. P. P. would like to acknowledge the Funda\c{c}\~{a}o de Amparo \`{a} Pesquisa e ao Desenvolvimento Cient\'{i}fico e Tecnol\'{o}gico do Maranh\~{a}o (FAPEMA).
\end{acknowledgments}

\setcounter{equation}{0}
\setcounter{table}{0}
\setcounter{section}{0}
\numberwithin{equation}{section}
\makeatletter
\renewcommand{\thesection}{\Alph{section}} 
\renewcommand{\thesubsection}{\Alph{section}.\arabic{subsection}}
\def\@gobbleappendixname#1\csname thesubsection\endcsname{\Alph{section}.\arabic{subsection}}
\renewcommand{\theequation}{\Alph{section}\arabic{equation}}
\renewcommand{\thefigure}{\arabic{figure}}
\renewcommand{\bibnumfmt}[1]{[#1]}
\renewcommand{\citenumfont}[1]{#1}

\section*{Appendix}


\section{Matrix powers of density matrices}
\label{sec:0000000000A}

In this Appendix, we address the proof of the integral representation of rational powers of density matrices as follows 
\begin{equation}
\label{eq:A000000001}
{\rho^s} = \frac{\sin(\pi{s})}{\pi}\,{\int_0^{\infty}}\,dx \, {x^{s - 1}} {\rho}{(\rho + x\mathbb{I})^{-1}} ~,
\end{equation}
with $0 < s < 1$, while $\mathcal{S} = \{\rho\in\mathcal{H} \mid {\rho^{\dagger}} = \rho,~\rho \geq 0,~\text{Tr}(\rho) = 1\}$ stands as a convex subspace of a $d$-dimensional Hilbert space $\mathcal{H}$, with $d = \dim\mathcal{H}$. To do so, let $\rho = {\sum_j}\,{p_j}|j\rangle\langle{j}|$ be a nonsingular, full rank density matrix, where $\{{p_j}\}_{j = 1,\ldots,d}$ and $\{|j\rangle\}_{j = 1,\ldots,d}$ define the sets of eigenvalues and eigenstates, respectively. We note that $0 < {p_j} < 1$ and ${\sum_j}\,{p_j} = 1$, $\langle{j}|{\ell}\rangle = {\delta_{j\ell}}$ and ${\sum_j}|j\rangle\langle{j}| = \mathbb{I}$, where $\mathbb{I}$ is the identity matrix. Hence, by exploiting the aforementioned spectral decomposition of the density matrix, one gets
\begin{equation}
\label{eq:A000000002}
{\rho^s} = {\sum_j} \,{p_j^s}|j\rangle\langle{j}| ~.
\end{equation}
We note that, for $s\in(0,1)$ and ${p_j} \in (0,1)$, one finds that $0 < {p_j^s} < 1$, for all $j \in \{1,\ldots,d\}$. It can be verified that, for $0 < s < 1$, the monotone function $p_j^s$ exhibits the following representation~\cite{Bathia}
\begin{equation}
\label{eq:A000000003}
{p_j^s} = \frac{\sin(\pi{s})}{\pi}\,{\int_0^{\infty}}\,dx \, {x^{s - 1}}{p_j}{({p_j} + x)^{-1}} ~.
\end{equation}
Hence, by combining Eqs.~\eqref{eq:A000000002} and~\eqref{eq:A000000003}, it yields
\begin{align}
\label{eq:A000000004}
{\rho^s} &= \frac{\sin(\pi{s})}{\pi}\,{\int_0^{\infty}}\,dx \, {x^{s - 1}} {\sum_j} \, {p_j}{({p_j} + x)^{-1}} |j\rangle\langle{j}| \nonumber\\
&= \frac{\sin(\pi{s})}{\pi}\,{\int_0^{\infty}}\,dx \, {x^{s - 1}} {\sum_{j,\ell}} \, {p_j}{({p_{\ell}} + x)^{-1}} |j\rangle\langle{j}|\ell\rangle\langle\ell| \nonumber\\
&= \frac{\sin(\pi{s})}{\pi}\,{\int_0^{\infty}}\,dx \, {x^{s - 1}} \left({\sum_j} \, {p_j} |j\rangle\langle{j}|\right) \nonumber\\
&\times\left({\sum_{\ell}}\, {({p_{\ell}} + x)^{-1}}|\ell\rangle\langle\ell|\right) ~, \nonumber\\
\end{align}
where we have used the orthogonality constraint $\langle{j}|\ell\rangle = {\delta_{j\ell}}$ respective to the eigenstates of the density matrix. Next, by using the completeness relation ${\sum_{\ell}}|\ell\rangle\langle{\ell}| = \mathbb{I}$, one finds
\begin{align}
\label{eq:A000000005}
{(\rho + x\mathbb{I})^{-1}} &= {\left({\sum_{\ell}}\, ({p_{\ell}} + x)|\ell\rangle\langle\ell|\right)^{-1}} \nonumber\\
&=  {\sum_{\ell}}{({p_{\ell}} + x)^{-1}}|\ell\rangle\langle\ell| ~.
\end{align}
In other words, Eq.~\eqref{eq:A000000005} means that the inverse matrix of a given nonsingular operator can be evaluated in terms of its spectral decomposition, taking into account the inverse eigenvalues accordingly.

Finally, combining Eqs.~\eqref{eq:A000000004} and~\eqref{eq:A000000005}, also recognizing the spectral decomposition of the density matrix, we conclude the integral representation of powers of density matrices given in Eq.~\eqref{eq:A000000001}. Importantly, we note that the nonsingular operators $\rho$ and ${(\rho + x\mathbb{I})^{-1}}$ commute with each other, i.e.,
\begin{align}
\label{eq:A000000006}
{\rho}{(\rho + x\mathbb{I})^{-1}} &= {[(\rho + x\mathbb{I}){\rho^{-1}}]^{-1}} \nonumber\\
&= {[{\rho^{-1}}(\rho + x\mathbb{I})]^{-1}} \nonumber\\
&= {(\rho + x\mathbb{I})^{-1}}\rho ~,
\end{align}
where we have used that $\rho{\rho^{-1}} = {\rho^{-1}}\rho$. This implies that Eq.~\eqref{eq:A000000001} can also be cast as follows:
\begin{align}
\label{eq:A000000007}
{\rho^s} &= \frac{\sin(\pi{s})}{\pi}\,{\int_0^{\infty}}\,dx \, {x^{s - 1}} {(\rho + x\mathbb{I})^{-1}} {\rho} ~.
\end{align}
We note that Eqs.~\eqref{eq:A000000001} and~\eqref{eq:A000000007} also hold when assuming the generalized inverse of operators, i.e., the inverse of the density matrix is taken on the set of eigenstates with nonzero eigenvalues.


\section{Bounding $\alpha$-$z$-RRE}
\label{sec:0000000000B}
 
In this Appendix, we present details on the derivation of the bounds for non-symmetric and symmetric $\alpha$-$z$-RREs discussed in the main text, namely Eqs.~\eqref{eq:0000000000004},~\eqref{eq:0000000000006}, and~\eqref{eq:0000000000007}. Let ${\rho_0},{\rho_t} \in \mathcal{S}$ be full rank, invertible density matrices of a finite-dimensional quantum system that undergoes a general physical process described by the map $\mathcal{E}(\bullet)$. In this setting, we have that ${\rho_0}$ stands as its initial state, while ${\rho_t} = {\mathcal{E}_t}({\rho_0})$ describes the respective instantaneous state, for $t \geq 0$. Hereafter, we consider $\alpha$-$z$-RRE evaluated over the range of parameters $0 < \alpha < 1$ and $1 \geq z \geq \text{max}\{\alpha, 1 - \alpha\}$.


\subsection{Bounding $\boldsymbol{{\text{D}_{\alpha,z}}({\rho_{\tau}}\|{\rho_0})}$}
\label{sec:0000000000B01}

We start by proving the result in Eq.~\eqref{eq:0000000000004}. To do so, we note that the absolute value of the time derivative of $\alpha$-$z$-RRE [see Eq.~\eqref{eq:0000000000001}] is written as
\begin{equation}
\label{eq:B000000001}
\left|\frac{d}{dt}{\text{D}_{\alpha,z}}({\rho_t}\|{\rho_0})\right| = \frac{[{g_{\alpha,z}}({\rho_t},{\rho_0})]^{-1}}{|1 - \alpha|}\left|\frac{d}{dt}{g_{\alpha,z}}({\rho_t},{\rho_0})\right| ~,
\end{equation}
where the $\alpha$-$z$-relative purity of states $\rho_0$ and $\rho_t$ is given by
\begin{equation}
\label{eq:B000000002}
{g_{\alpha,z}}({\rho_t},{\rho_0}) = \text{Tr}\left[{\left({\rho_0^{\frac{1 - \alpha}{2z}}}{\rho_t^{\frac{\alpha}{z}}}{\rho_0^{\frac{1 - \alpha}{2z}}}\right)^z}\right] ~.
\end{equation}
We remind the reader that, for $\alpha \in (0,1)$ and $z > 0$, one finds ${g_{\alpha,z}}({\rho_t},{\rho_0}) \leq 1$, while ${g_{\alpha,z}}({\rho_t},{\rho_0}) = 1$, if and only if ${\rho_t} = {\rho_0}$ for a given $t \geq 0$~\cite{ALPHAzAuden,DPI_ZHANG001}. It is worth noting that $\alpha$-$z$-relative purity satisfies the inequality
\begin{equation}
\label{eq:B000000003}
{\left[{g_{\alpha,z}}({\rho_t},{\rho_0})\right]^{-1}} \leq {\left\{1 + (1 - \alpha)\ln[{k_{\text{min}}}({\rho_0})]\right\}^{-1}} ~.
\end{equation}
To see this, one considers the Araki-Lieb-Thirring inequality, $\text{Tr}[{({A^r}{B^r}{A^r})^q}] \geq \text{Tr}[{(ABA)^{qr}}]$, which holds for real-valued parameters $q \geq 0$ and $r \geq 1$~\cite{araki,Lieb2002}. Hence, by choosing the Hermitian operators $A = {\rho_0^{(1 - \alpha)/2}}$ and $B = {\rho_t^{\alpha}}$, with $q = z$ and $r = 1/z$, we find the lower bound
\begin{equation}
\label{eq:B000000004}
{g_{\alpha,z}}({\rho_t},{\rho_0}) \geq \text{Tr}\left({\rho_t^{\alpha}}{\rho_0^{1 - \alpha}}\right) ~,
\end{equation}
where we have used the cyclic property of trace. In turn, for $\alpha \in (0,1)$, the quantity on the right-hand side of the inequality in Eq.~\eqref{eq:B000000004} was proven to be bounded from below as~\cite{bounding_generalized}
\begin{equation}
\label{eq:B000000005}
\text{Tr}({\rho_t^{\alpha}}{\rho_0^{1 - \alpha}}) \geq 1 + (1 - \alpha)\ln[{k_{\text{min}}}({\rho_0})] ~.
\end{equation}
Therefore, by combining Eqs.~\eqref{eq:B000000004} and~\eqref{eq:B000000005}, one readily concludes the aforementioned result in Eq.~\eqref{eq:B000000003}.

Next, we find that the absolute value of the rate of change of $\alpha$-$z$-relative purity fulfills the upper bound as follows
\begin{align}
\label{eq:B000000006}
&\left|\frac{d}{dt}{g_{\alpha,z}}({\rho_t},{\rho_0})\right| \leq \nonumber\\
&\alpha\left[\frac{k_{\text{max}}({\rho_0})}{\left({k_{\text{min}}}({\rho_0})\right)^{1 - z}\left({k_{\text{min}}}({\rho_t})\right)^z}\right]^{\frac{1 - \alpha}{z}}{\left\|\frac{d{\rho_t}}{dt}\right\|_1} ~.
\end{align}
To prove the latter result, we first note that the absolute value of the time derivative of Eq.~\eqref{eq:B000000002} is written as
\begin{align}
\label{eq:B000000007}
&\left|\frac{d}{dt}{g_{\alpha,z}}({\rho_t},{\rho_0})\right| =\nonumber\\
& z \left|\text{Tr}\left({\rho_0^{\frac{1 - \alpha}{2z}}}{\left({\rho_0^{\frac{1 - \alpha}{2z}}}{\rho_t^{\frac{\alpha}{z}}}{\rho_0^{\frac{1 - \alpha}{2z}}}\right)^{z - 1}}{\rho_0^{\frac{1 - \alpha}{2z}}} \, \frac{d{\rho_t^{\frac{\alpha}{z}}}}{dt}\right)\right| ~,
\end{align}
where we have used the algebraic relation $d \text{Tr}({w_t^z})/dt = z\text{Tr}[{w_t^{z - 1}}(d{w_t}/dt)]$, and also the cyclic property of trace, with ${w_t} := {\rho_0^{(1 - \alpha)/2z}}{\rho_t^{\alpha/z}}{\rho_0^{(1 - \alpha)/2z}}$ being a time-dependent, full-rank, Hermitian operator. This identity can be verified straightforwardly by exploiting the spectral decomposition of operator $w_t$, and it will be valid for all $z$. Second, we have that operator $\rho_t^{\alpha/z}$ can be cast in terms of its integral representations as
\begin{equation}
\label{eq:B000000008}
{\rho_t^{\frac{\alpha}{z}}} = \frac{\sin(\alpha\pi/z)}{\pi}{\int_0^{\infty}} dx \, {x^{\frac{\alpha}{z} - 1}}{\rho_t}{({\rho_t} + x\mathbb{I})^{-1}} ~,
\end{equation}
with $\mathbb{I}$ being the identity matrix, whenever $0 < \alpha/z < 1$. For details on Eq.~\eqref{eq:B000000008}, see Appendix~\ref{sec:0000000000A}. This condition is in accordance with the chosen parameter range $0 < \alpha < 1$ and $1 \geq z \geq \text{max}\{\alpha, 1 - \alpha\}$. In this case, the time derivative of such an operator reads
\begin{align}
\label{eq:B000000009}
&\frac{d{\rho_t^{\frac{\alpha}{z}}}}{dt} = \nonumber\\
& \frac{\sin(\alpha\pi/z)}{\pi}{\int_0^{\infty}} dx \, {x^{\frac{\alpha}{z}}}  {({\rho_t} + x\mathbb{I})^{-1}}\frac{d{\rho_t}}{dt}{({\rho_t} + x\mathbb{I})^{-1}} ~,
\end{align}
where we have used the algebraic identity ${\rho_t}{({\rho_t} + x\mathbb{I})^{-1}} = \mathbb{I} - x{({\rho_t} + x\mathbb{I})^{-1}}$, and also the relation $d{({\rho_t} + x\mathbb{I})^{-1}}/dt = - {({\rho_t} + x\mathbb{I})^{-1}}(d{\rho_t}/dt){({\rho_t} + x\mathbb{I})^{-1}}$. We now combine Eqs.~\eqref{eq:B000000007},~\eqref{eq:B000000008}, and~\eqref{eq:B000000009}, thus obtaining the result
\begin{align}
\label{eq:B000000010}
&\left|\frac{d}{dt}{g_{\alpha,z}}({\rho_t},{\rho_0})\right| = \frac{z\sin(\alpha\pi/z)}{\pi} \left|{\int_0^{\infty}} dx \, {x^{\frac{\alpha}{z}}} \text{Tr}\left({({\rho_t} + x\mathbb{I})^{-1}} \right.\right. \nonumber\\
&\left.\left.\times {\rho_0^{\frac{1 - \alpha}{2z}}} {\left({\rho_0^{\frac{1 - \alpha}{2z}}} {\rho_t^{\frac{\alpha}{z}}} {\rho_0^{\frac{1 - \alpha}{2z}}}\right)^{z - 1}} {\rho_0^{\frac{1 - \alpha}{2z}}} {({\rho_t} + x\mathbb{I})^{-1}} \frac{d{\rho_t}}{dt} \right)\right| ~,
\end{align}
where we have used the cyclic property of the trace operation. Next, applying the triangle inequality $|\int du \, f(u)| \leq \int du \, |f(u)|$ to the right-hand side of Eq.~\eqref{eq:B000000010}, we verify that the absolute value of the time derivative of $\alpha$-$z$-relative purity satisfies the following upper bound
\begin{align}
\label{eq:B000000011}
&\left|\frac{d}{dt}{g_{\alpha,z}}({\rho_t},{\rho_0})\right| \leq \frac{z\sin(\alpha\pi/z)}{\pi}{\int_0^{\infty}} dx \, {x^{\frac{\alpha}{z}}} \left|\text{Tr}\left({({\rho_t} + x\mathbb{I})^{-1}} \right.\right.\nonumber\\
&\left.\left. \times{\rho_0^{\frac{1 - \alpha}{2z}}} {\left({\rho_0^{\frac{1 - \alpha}{2z}}} {\rho_t^{\frac{\alpha}{z}}} {\rho_0^{\frac{1 - \alpha}{2z}}}\right)^{z - 1}} {\rho_0^{\frac{1 - \alpha}{2z}}} {({\rho_t} + x\mathbb{I})^{-1}} \frac{d{\rho_t}}{dt} \right)\right| ~.
\end{align}

It is known that the absolute value of the trace of the product of a given set of operators satisfies the inequalities 
\begin{equation}
\label{eq:B000000012}
\left|\text{Tr}\left({\prod_{l = 1}^6}\,{A_l}\right)\right| \leq {\left\|{\prod_{l = 1}^5}\,{A_l}\right\|_{\infty}}{\|{A_6}\|_1} \leq {\prod_{l = 1}^5}\,{\|{A_l}\|_{\infty}}{\|{A_6}\|_1} ~.
\end{equation}
In detail, the first inequality comes from H\"{o}lder's ine\-quality for Schatten norms, $|\text{Tr}({A_r^{\dagger}}{A_s})| \leq {\|{A_r}\|_p}{\|{A_s}\|_q}$, with $1/p + 1/q = 1$, and general operators $A_r$ and $A_s$. In particular, by fixing $p = 1$ and $q = \infty$, one finds that the absolute value of a trace of operators satisfies the upper bound $|\text{Tr}({A_r^{\dagger}}{A_s})| \leq {\|{A_r}\|_{\infty}}{\|{A_s}\|_1}$~\cite{Bathia}. The second inequality comes from the fact that the Schatten $p$-norm is sub-multiplicative, ${\|{A_r}{A_s}\|_p} \leq {\|{A_r}\|_p}{\|{A_s}\|_p}$, with $p \in [1,\infty)$~\cite{Watrous}. In words, we have that the product of two operators is bounded from above by the product of their individual Schatten norms. Next, we combine Eqs.~\eqref{eq:B000000011} and~\eqref{eq:B000000012}, with ${A_1} = {A_5} = {({\rho_t} + x\mathbb{I})^{-1}}$, ${A_2} = {A_4} = {\rho_0^{\frac{1 - \alpha}{2z}}}$, ${A_3} = {\left({\rho_0^{\frac{1 - \alpha}{2z}}} {\rho_t^{\frac{\alpha}{z}}} {\rho_0^{\frac{1 - \alpha}{2z}}}\right)^{z - 1}}$, and also ${A_6} = d{\rho_t}/dt$, and we conclude
\begin{align}
\label{eq:B000000013}
&\left|\frac{d}{dt}{g_{\alpha,z}}({\rho_t},{\rho_0})\right| \leq {\left\|{\rho_0^{\frac{1 - \alpha}{2z}}}\right\|_{\infty}^2} {\left\|{\left({\rho_0^{\frac{1 - \alpha}{2z}}}{\rho_t^{\frac{\alpha}{z}}}{\rho_0^{\frac{1 - \alpha}{2z}}}\right)^{z - 1}}\right\|_{\infty}} \nonumber\\
 &\times {\left\|\frac{d{\rho_t}}{dt}\right\|_1} \, \frac{z\sin(\alpha\pi/z)}{\pi} {\int_0^{\infty}} dx \frac{x^{\frac{\alpha}{z}}}{({k_{\text{min}}}({\rho_t}) + x)^2} ~,
\end{align}
where we have used that ${\|{({\rho_t} + x\mathbb{I})^{-1}}\|_{\infty}} = {\left({k_{\text{min}}}({\rho_t}) + x\right)^{-1}}$, and ${k_{\text{min}/\text{max}}}(\bullet)$ sets the smal\-lest/largest eigenvalue. In turn, given the a-na\-ly\-ti\-cal result ${\int_0^{\infty}} dx \, {x^{\alpha/z}}{\left({k_{\text{min}}}({\rho_t}) + x\right)^{-2}} = (\alpha\pi/z)({k_{\text{min}}}({\rho_t}))^{\alpha/z - 1} \csc(\alpha\pi/z)$, Eq.~\eqref{eq:B000000013} becomes
\begin{align}
\label{eq:B000000014}
&\left|\frac{d}{dt}{g_{\alpha,z}}({\rho_t},{\rho_0})\right| \leq \alpha {\left\|{\rho_0^{\frac{1 - \alpha}{2z}}}\right\|_{\infty}^2} {\left\|{\left({\rho_0^{\frac{1 - \alpha}{2z}}}{\rho_t^{\frac{\alpha}{z}}}{\rho_0^{\frac{1 - \alpha}{2z}}}\right)^{z - 1}}\right\|_{\infty}} \nonumber\\
&\times {\left({k_{\text{min}}}({\rho_t})\right)^{\frac{\alpha}{z} - 1}} {\left\|\frac{d{\rho_t}}{dt}\right\|_1} ~.
\end{align}
The operator norm satisfies the upper bound ${\|{A_1^p} \, {A_2^q}\|_{\infty}} \leq {\|{A_1^p}\|_{\infty}} {\|{A_2^q}\|_{\infty}}$, with ${\|{A_l^p}\|_{\infty}} = {\|{A_l}\|^p_{\infty}}$, for all $p,q \in (0,1)$, $l = \{1,2\}$. We also note the algebraic relation ${({A^a})^b} = {({A^b})^a}$, for all $a, b \in \mathbb{R}$, which in turn can be verified using the spectral decomposition of some normal operator $A$. Hence, we arrive at the following inequalities 
\begin{equation}
\label{eq:B000000015}
{\left\|{\rho_0^{\frac{1 - \alpha}{2z}}}\right\|_{\infty}^2} \leq {\|{\rho_0}\|_{\infty}^{\frac{1 - \alpha}{z}}} ~,
\end{equation}
and
\begin{align}
\label{eq:B000000016}
&{\left\|{\left({\rho_0^{\frac{1 - \alpha}{2z}}}{\rho_t^{\frac{\alpha}{z}}}{\rho_0^{\frac{1 - \alpha}{2z}}}\right)^{z - 1}}\right\|_{\infty}} \nonumber\\
&= {\left\|{\left({\left({\rho_0^{\frac{1 - \alpha}{2z}}}\right)^{-1}}{\left({\rho_t^{\frac{\alpha}{z}}}\right)^{-1}}{\left({\rho_0^{\frac{1 - \alpha}{2z}}}\right)^{-1}}\right)^{1 - z}}\right\|_{\infty}} \nonumber\\
&\leq {\left\|{\left({\rho_0^{-1}}\right)^{\frac{1 - \alpha}{2z}}}{\left({\rho_t^{-1}}\right)^{\frac{\alpha}{z}}}{\left({\rho_0^{-1}}\right)^{\frac{1 - \alpha}{2z}}}\right\|^{1 - z}_{\infty}} \nonumber\\
&\leq \left\Vert \rho_{0}^{-1}\right\Vert _{\infty}^{\frac{(1-\alpha)(1-z)}{z}}\left\Vert \rho_{t}^{-1}\right\Vert _{\infty}^{\frac{\alpha(1 - z)}{z}} ~.
\end{align}
By combining Eqs.~\eqref{eq:B000000014},~\eqref{eq:B000000015} and~\eqref{eq:B000000016}, we obtain the upper bound indicated in Eq.~\eqref{eq:B000000006}. Next, by substituting Eqs.~\eqref{eq:B000000003} and \eqref{eq:B000000006} into Eq.~\eqref{eq:B000000001}, we conclude
\begin{equation}
\label{eq:B000000017}
\left|\frac{d}{dt}{\text{D}_{\alpha,z}}({\rho_t}\|{\rho_0})\right| \leq \frac{\alpha \, {h_{\alpha,z}}({\rho_0})}{\left|1 - \alpha\right|} {\left[{k_{\text{min}}}({\rho_t})\right]^{\alpha - 1}}{\left\|\frac{d{\rho_t}}{dt}\right\|_1} ~,
\end{equation}
where
\begin{equation}
\label{eq:B000000018}
{h_{\alpha,z}}({\rho_0}) := \frac{\left({k_{\text{max}}}({\rho_0})\left[{k_{\text{min}}}({\rho_0})\right]^{z - 1}\right)^{\frac{1 - \alpha}{z}}}{\left|1 + (1 - \alpha)\ln[k_{\text{min}}({\rho_0})]\right|} ~.
\end{equation}
Finally, integrating Eq.~\eqref{eq:B000000017} over the time interval $t \in [0,\tau]$, and also using the triangle inequality $|\int du \, f(u)| \leq \int du \, |f(u)|$, we conclude the result given in Eq.~\eqref{eq:0000000000004}.


\subsection{Bounding $\boldsymbol{{\text{D}_{\alpha,z}}({\rho_0}\|{\rho_{\tau}})}$}
\label{sec:0000000000B02}
 
Here, we prove the result in Eq.~\eqref{eq:0000000000006}. To do so, we explore the fact that $\alpha$-$z$-RRE is skew symmetric, ${\text{D}_{\alpha,z}}({\rho_0}\|{\rho_t}) = [{\alpha}/{(1 - \alpha)}]{\text{D}_{1 - \alpha,z}}({\rho_t}\|{\rho_0})$, for all $\alpha \in (0,1)$ [see Sec.~\ref{sec:00000000002}]. Next, by taking the absolute value of the time derivative of this result, one finds
\begin{equation}
\label{eq:B000000019}
\left|\frac{d}{dt}{\text{D}_{\alpha,z}}({\rho_0}\|{\rho_t})\right| = \frac{\alpha}{|1 - \alpha|}\left|\frac{d}{dt}{\text{D}_{1 - \alpha,z}}({\rho_t}\|{\rho_0})\right| ~.
\end{equation}
We now evaluate Eq.~\eqref{eq:B000000017} under the mapping $\alpha \rightarrow 1 - \alpha$, and we combine the resulting expression with Eq.~\eqref{eq:B000000019}, which implies the upper bound
\begin{equation}
\label{eq:B000000020}
\left|\frac{d}{dt}{\text{D}_{\alpha,z}}({\rho_0}\|{\rho_t})\right| \leq {h_{1 - \alpha, z}}({\rho_0}){\left[{k_{\text{min}}}({\rho_t})\right]^{-\alpha}}{\left\|\frac{d{\rho_t}}{dt}\right\|_1} ~,
\end{equation}
where the auxiliary function ${h_{1 - \alpha, z}}({\rho_0})$ is obtained from Eq.~\eqref{eq:B000000018}. In conclusion, we integrate Eq.~\eqref{eq:B000000020} over the time interval $0 \leq t \leq \tau$, apply the triangle inequality $|\int du \, f(u)| \leq \int du \, |f(u)|$, and thus we arrive at Eq.~\eqref{eq:0000000000006} of the main text.


\subsection{Bounding $\boldsymbol{{\text{D}_{\alpha,z}}({\rho_0}:{\rho_{\tau}})}$}
\label{sec:0000000000B03}

Finally, we comment on the proof of Eq.~\eqref{eq:0000000000007}. Starting from the symmetric $\alpha$-$z$-RRE evaluated to the states $\rho_0$ and $\rho_t$ in Eq.~\eqref{eq:0000000000003}, the absolute value of its time derivative yields
\begin{align}
\label{eq:B000000021}
&\left|\frac{d}{dt}{\text{D}_{\alpha,z}}({\rho_0}:{\rho_t})\right| \leq \left|\frac{d}{dt}{\text{D}_{\alpha,z}}({\rho_t}\|{\rho_0})\right| + \left|\frac{d}{dt}{\text{D}_{\alpha,z}}({\rho_0}\|{\rho_t})\right| ~,
\end{align}
where we have used the triangle inequality $|a + b| \leq |a| + |b|$. Next, substituting Eqs.~\eqref{eq:B000000017} and~\eqref{eq:B000000020} into Eq.~\eqref{eq:B000000021}, we obtain the upper bound as follows
\begin{align}
\label{eq:B000000022}
&\left|\frac{d}{dt}{\text{D}_{\alpha,z}}({\rho_0}:{\rho_t})\right| \leq \frac{\alpha \, {h_{\alpha,z}}({\rho_0})}{\left|1 - \alpha\right|} {\left[{k_{\text{min}}}({\rho_t})\right]^{\alpha - 1}}{\left\|\frac{d{\rho_t}}{dt}\right\|_1} \nonumber\\
&+ {h_{1 - \alpha, z}}({\rho_0}){\left[{k_{\text{min}}}({\rho_t})\right]^{-\alpha}}{\left\|\frac{d{\rho_t}}{dt}\right\|_1} ~.
\end{align}
To conclude the proof, we integrate Eq.~\eqref{eq:B000000022} over the time interval $t \in [0,\tau]$, and we apply the triangle inequality $|\int du \, f(u)| \leq \int du \, |f(u)|$. Therefore, we explicitly check Eq.~\eqref{eq:0000000000007}.



%

\end{document}